\documentclass[fleqn,usenatbib]{mnras}

\usepackage{newtxtext}
\usepackage{graphicx}	
\usepackage{amsmath}	
\usepackage{amssymb}	
\usepackage{bm}
\usepackage{breakcites}
\usepackage{url} 
\usepackage{float}
\usepackage{subfigure}
\usepackage{color}
\usepackage{lineno} 
\usepackage{enumitem}  
\usepackage{chngcntr}
\usepackage{cancel}
\usepackage{threeparttable}
\usepackage{tikz}
\usepackage{academicons}
\usepackage{adjustbox}
\usepackage[T1]{fontenc}

\usepackage{graphicx}	
\usepackage{amsmath}	
\usepackage{amssymb}	
\usepackage{hyperref}
\usepackage{xspace}

\definecolor{frenchblue}{rgb}{0.0, 0.45, 0.73}
\definecolor{burgundy}{rgb}{0.5, 0.0, 0.13}
\definecolor{royalblue}{RGB}{65,105,225}
\definecolor{darkspringgreen}{rgb}{0.09, 0.45, 0.27}

\graphicspath{{./}{plots/}{figures/}}

\definecolor{lime}{HTML}{A6CE39}
\DeclareRobustCommand{\orcidicon}{%
	\begin{tikzpicture}
	\draw[lime, fill=lime] (0,0) 
	circle [radius=0.16] 
	node[white] {{\fontfamily{qag}\selectfont \tiny ID}};
	\draw[white, fill=white] (-0.0625,0.095) 
	circle [radius=0.007];
	\end{tikzpicture}
	\hspace{-2mm}
}

\foreach \x in {A, ..., Z}{%
	\expandafter\xdef\csname orcid\x\endcsname{\noexpand\href{https://orcid.org/\csname orcidauthor\x\endcsname}{\noexpand\orcidicon}}
}

\newcommand{\orcid}[1]{\href{https://orcid.org/#1}{\textcolor[HTML]{A6CE39}{\orcidicon}}}

\newcommand{\ltsima}{$\buildrel < \over \sim$}
\newcommand{\lsim}{\lower.5ex\hbox{\ltsima}}
\newcommand{\gtsima}{$\buildrel > \over \sim$}
\newcommand{\gsim}{\lower.5ex\hbox{\gtsima}}

\newcommand{\sj}{\hbox{Swift\,J0243.6+6124}\xspace}
\newcommand{\sjs}{\hbox{J0243}\xspace}

\title[Radio from colliding winds in NS XRBs]{Radio emission from colliding outflows in high-mass X-ray binaries with strongly magnetized neutron stars}

\author[M.~Chatzis et al.]{
Margaritis~Chatzis,$^{1}$\thanks{E-mail: ph5123@edu.physics.uoc.gr}
Maria~Petropoulou\orcid{0000-0001-6640-0179},$^{2}$\thanks{E-mail: mpetropo@phys.uoa.gr}
Georgios~Vasilopoulos\orcid{0000-0003-3902-3915}$^{3,4}$
\\
$^{1}$Department of Physics, University of Crete, GR-71003, Heraklion, Greece\\
$^{2}$Department of Physics, National and Kapodistrian University of Athens, University Campus Zografos, GR 15783, Athens, Greece\\
$^{3}$Université de Strasbourg, CNRS, Observatoire astronomique de Strasbourg, Strasbourg, 67000, France\\
$^{4}$Department of Astronomy, Yale University, PO Box 208101, New Haven, CT 06520-8101, USA\\
}

\date{Accepted XXX. Received YYY; in original form ZZZ}

\pubyear{2021}

\begin{document}
\label{firstpage}
\pagerange{\pageref{firstpage}--\pageref{lastpage}}
\maketitle

\begin{abstract}

We present a toy model for radio emission in HMXBs with strongly magnetized NSs where a wind-collision region is formed by the NS outflow and the stellar wind of the massive companion.  Radio emission is expected from the synchrotron radiation of shock-accelerated electrons  and  the free-free emission of the stellar wind. We found that the predicted relation between the GHz luminosity  ($L_R$) and the accretion X-ray luminosity ($L_X$) can be written as $L_R \propto L_X^b$ for most parameters. No correlation with X-rays is expected ($b=0$) when the thermal emission of the stellar wind dominates in radio. We typically find a steep correlation ($b=12/7$)  for sub-Eddington X-ray luminosities and a more shallow one ($b=2(p-1)/7$) for super-Eddington X-ray luminosities, where $p$ is the power-law index of accelerated electrons. The maximum predicted radio luminosity is independent of the NS properties, while it depends on the stellar wind momentum, binary separation distance, and the minimum electron Lorentz factor.  Using a Bayesian approach we modelled the radio observations of \sj that cover a wide range of mass accretion rates. Our results support a shock origin for the radio detections at sub-Eddington X-ray luminosities. However, no physically meaningful parameters could be found for the super-Eddington phase of the outburst, suggesting a different origin. Future observations with more sensitive instruments might reveal a large number of HMXBs with strongly magnetized NSs in radio, allowing determination of the slope in the $L_R-L_X$ relation, and putting the wind-collision scenario into test. 
\end{abstract}

\begin{keywords}
radiation mechanism: non-thermal, thermal -- stars: neutron  -- X-rays: binaries
\end{keywords}



\section{Introduction}

X-ray binaries (XRBs) are systems where mass is transferred from a donor star onto a compact object while its dynamical energy is converted to radiation. XRBs that host stellar mass black holes or neutron stars (NSs) with low magnetic fields (i.e. $<10^9$ G) tend to spent most of their lives in quiescence but occasionally undergo major outbursts that may reach or exceed the Eddington limit. Studies of the outburst/quiescence cycle of such systems have revealed dramatic changes in their X-ray spectral characteristics and temporal properties along with changes in their luminosity \citep[][]{2006csxs.book..157M}.
The main spectral states during outbursts are known as the hard state,  which is characterised by a Comptonised X-ray spectrum, and the soft state, which is dominated by thermal emission from an accretion disc.
Moreover, multi-wavelength observations have revealed the presence of radio emission associated with a compact jet during the hard and quiescent state, providing evidence for a correlation between the radio and X-ray fluxes~\citep[e.g.][]{1998A&A...337..460H, 2003A&A...397..645M, 2003MNRAS.344...60G, 2004ApJ...617.1272C,2013MNRAS.428.2500C}. 

In fact, the radio/X-ray correlation is a key piece of observational evidence about the disc-jet (or inflow-outflow) connection in XRBs.
The `universal' radio/X-ray flux correlation ($L_R \propto L_X^{0.6}$) seems to hold for most BH XRBs in low-hard flux states, although data are dominated by two systems, GX 339-4 and V404 Cyg~\citep[e.g.][]{2012MNRAS.423..590G, 2013MNRAS.428.2500C}, and more recently MAXI J1820+070 \citep{2020NatAs...4..697B}. This relation holds even for some of the faintest BH systems like A0620-00 \citep{2006MNRAS.370.1351G,  2018ApJ...852....4D} and XTE~J1118+480  \citep{2014MNRAS.445..290G}.
However, the radio/X-ray luminosity plane contains outliers like H~1743-322 \citep{2011IAUS..275..255C}, where the correlation is steeper ($L_R \propto L_X^{1.4}$) and quite similar to the one found in the few reported NS XRBs \citep[e.g.][]{2006MNRAS.366...79M}. 

Till recently radio emission associated with jets was thought to be a unique property of BH XRBs or those hosting  weakly magnetized NSs (i.e. $<10^9$ G). However, the narrative changed following deep observations with the Very Large
Array (VLA) telescope that led to the discovery of radio emission from the prototypical X-ray pulsar Her~X-1 \citep{2018MNRAS.473L.141V} and the symbiotic X-ray binary GX~1+4 \citep{2018MNRAS.474L..91V}. More recently, X-ray pulsars \sj and 1A~0535+262 were detected in radio during major outbursts \citep{2018Natur.562..233V, 2020ATel14193....1V}. Both of these systems are high-mass X-ray binaries (HMXBs) with a Be donor \citep[i.e. BeXRBs,][]{2011Ap&SS.332....1R}. BeXRBs are quite different than BH XRBs, for in BeXRBs material escapes the massive donor through a slow moving equatorial outflow, that is usually referred as decretion disc or Be disc \citep[e.g.][]{2011A&A...527A..84K}.
The disc itself is quite complex and can be described as a circumstellar environment with atomic, molecular, and dusty regions usually found in rings or disc-like structures \citep[e.g.][]{2018MNRAS.480..320M}.
BeXRBs are quite variable showing two types of outbursts. Type I outbursts ($L_X\sim10^{36}$~erg s$^{-1}$) may occur during a close passage of the NS to the decretion disc and thus show a correlation with the binary orbital period. Giant or Type II outbursts ($L_{X}\ge10^{38}$~erg s$^{-1}$) that can last for several orbits might be associated with wrapped Be-discs \citep{2013PASJ...65...41O}. During outbursts an accretion disc is formed around the NS, which  may be truncated at very large radii ($100-1000$~km) due to the large magnetic field of the NS (typically $>10^{12}$~G). Matter is then accreted following the magnetic field lines onto the NS surface forming a thermal mount of material, the so-called accretion column \citep{1976MNRAS.175..395B}. The accretion column is responsible for the bulk of the X-ray emission and any disc contribution is usually negligible at sub-Eddington luminosities\footnote{The Eddington luminosity is defined as $L_{\rm Edd}\simeq 1.5 \times10^{38} \, M_{\rm NS}/ M_{\sun}$~erg s$^{-1}$.}. 

The physical interpretation of the radio/X-ray correlation in BH XRBs has been widely discussed in the literature. While it is well established that during the low-hard states radio is produced by synchrotron-emitting electrons in a compact jet, the origin of the X-ray emission is less clear; X-rays could be part of the optically thin jet-synchrotron spectrum \citep[e.g.][]{2001A&A...372L..25M, 2003A&A...397..645M,2003MNRAS.345.1057M, 2004A&A...414..895F} or they could arise from a radiatively inefficient accretion flow or a radiatively efficient disc-corona system~\citep[e.g.][]{2003MNRAS.343L..59H, 2003MNRAS.345.1057M, 2005MNRAS.360L..68M, 2005ApJ...620..905Y}. In the case of highly magnetized X-ray pulsars, understanding the coupling of radio and X-ray emissions can be even more complicated. On the one hand, no jet structure has been resolved in any of the radio-detected accreting X-ray pulsars, so even the origin of the radio emission is less certain. On the other hand, the detected X-ray emission is mainly produced in the accretion column, close to the NS surface. Hence, even if the radio emission had a jet origin in X-ray pulsars, X-rays would still be produced in another region with different physical conditions, such as plasma densities and magnetic fields. In the latter case, deriving a relation coupling the X-ray and radio luminosities would require theoretical understanding of the jet formation in highly magnetized accreting pulsars, which is far from being settled.
Further observational studies of the $L_R$ and $L_X$ evolution during outbursts as well as in low flux states should be tested against theoretical predictions to better understand the physical connection behind this phenomenon.   

\subsection*{The case of Swift J0243.6+6124}

For Her X-1 it was originally proposed that the radio could be due to a synchrotron emitting radio jet \citep{2018MNRAS.473L.141V}. 
\sj (hereafter \sjs) was the first Galactic source that exceeded by a factor of 10 the Eddington limit for a NS \citep{2018ApJ...863....9W}, and was thus classified as a transient ultra-luminous X-ray pulsar \citep[ULXPs, see review][]{2017ARA&A..55..303K}. Its high luminosity and proximity made the source an ideal target for multi-wavelength monitoring, resulting in the first detection of radio emission from a BeXRB during an outburst. 
The radio jet scenario from an X-ray pulsar was reinforced following the initial discovery of correlated radio and X-ray emission from \sjs during the super-Eddington phase of the 2017-2018 outburst \citep{2018Natur.562..233V}.
However, extended monitoring of \sjs during the low flux state revealed a quite puzzling behaviour between radio and X-ray emission \citep{2019MNRAS.483.4628V}, with bright radio states as the source returned to a low X-ray flux state showing moderate type-I X-ray outbursts.

A solution proposed to explain these radio detections is based on the notion that a jet quickly re-establishes by driving the magnetospheric radius inward and outward with respect to the co-rotation radius when a certain X-ray luminosity level is crossed.  However, such a hypothesis has certain caveats. The timescale for the reappearance of the jet is remarkably shorter than in other X-ray binaries, taking only some days instead of the usual one to a few weeks \citep[e.g.][]{2013MNRAS.431L.107C,2013MNRAS.436.2625V}. Moreover, the launching site of the jet is not well understood \citep[e.g.][]{2008A&A...477....1M, 2016ApJ...822...33P} while all further calculations require an a priori assumption that the launch originates within the truncated disc~\citep[e.g.][]{2018Natur.562..233V}. 

Another possibility to explain the radio emission is to consider that accreting HMXBs can launch outflows.
These outflows can be launched during phases characterized by high accretion as those encountered in HMXB outbursts or in persistent ULXs \citep[see e.g.][for some theoretical predictions]{1973A&A....24..337S,2021arXiv211008249M}.
State-of-the-art numerical simulations also show the formation of outflows in super-Eddington ULXPs \citep[e.g.][]{2017ApJ...845L...9T,2021ApJ...917L..31A}, and in low-luminosity accreting systems \citep[e.g.][]{2017ApJ...851L..34P}.
Moreover, absorption features consistent with fast relativistic outflows have been observed in spectral studies of ULXPs \citep[e.g.][]{2018MNRAS.479.3978K,2019MNRAS.487.4355V}, or have been indirectly probed in order to explain their overall temporal properties \citep[e.g.][]{2019MNRAS.484..687M,2019MNRAS.488.5225V,2020MNRAS.491.4949V}.
%
%
%
These outflows can collide with wind from the donor star creating a shock-accelerated population of relativistic electrons that emits in radio frequencies. 
This thought, that colliding winds can produce a substantial amount of radio emission, is not unfounded. The radio emission of a plethora of massive binaries (e.g. Wolf Rayet-OB stars) can be explained by synchrotron emission originating in the wind-collision region \citep[e.g.][]{1986ApJ...303..239A, 1993ApJ...402..271E, 2003A&A...409..217D, 2006A&A...446.1001P,2013A&A...558A..28D}. The possibility that the radio emission comes from shocks between the accretion disc wind and the wind of a Wolf-Rayet donor star has also been discussed in the context of the microquasar Cygnus X-3 in its hypersoft X-ray state~\citep{2018A&A...612A..27K}.

Motivated by the above we will explore a scenario where radio emission is produced in the collision region of the NS outflow and the Be wind. We will build a toy model for computing synchrotron emission from shock-accelerated electrons taking into account synchrotron self-absorption, free-free absorption by the stellar wind, and Razin suppression due to background plasma. The contribution of the Be wind in radio frequencies due to free-free emission will also be considered. Our goal is to check if substantial radio emission can be produced from the system of colliding outflow-wind, and if so, what is the predicted relation between radio and X-ray luminosities. As a test case for our model we will use \sjs, for it is the only system with multiple radio detections that map a wide range of mass accretion rates.

This paper is structured as follows. In Section~\ref{sec:model} we outline the model used to compute the radio emission and present our results in Section~\ref{sec:results}. In Section~\ref{sec:application} we apply our model to the multi-epoch radio observations of \sjs. We continue with a discussion of our results and model caveats in Section~\ref{sec:discuss}. We conclude with a summary of our main findings in Section~\ref{sec:conclusions}.

\section{Model description}\label{sec:model}
In this section we will describe our toy model for the calculation of radio emission from HMXBs with highly magnetized NSs, while highlighting the main assumptions entering our calculations. We will first describe the assumed geometry of the system  (Sec.~\ref{sec:geometry}) and then the radiative processes considered (Sec.~\ref{sec:radiation}).

\subsection{A system of colliding outflows}\label{sec:geometry}

We consider a binary system composed of a NS and a Be star. The companion, being a massive early type star, loses mass at much higher rates than a solar-like star. Be stars are known to lose mass predominantly through equatorial winds, but they also possess an extended wind region composed of low density fast-moving plasma \citep[for a review, see][]{2003PASP..115.1153P}. Evidence for such anisotropic winds have been recently reported in interacting wind systems through X-ray observations~\citep[e.g.][]{2018MNRAS.474L..22P,2020MNRAS.495..365C}.
In our toy model we assume, however, a spherically symmetric (non-clumpy) wind with constant mass loss rate $\dot{M}_{\rm Be}$. We do not consider the acceleration of the Be wind in our calculations. Instead, we assume that the wind is coasting with its terminal velocity, $V_{\rm Be}^\infty$. The mass-loss rate of early type stars in the main sequence strongly depends on the stellar effective temperature. For instance, the hottest B stars ($T_{\rm eff}\simeq 3\times10^4$~K)  with masses $M \sim  12  M_{\odot}$  have $\dot{M}_{\rm Be} \sim 10^{-9} \, M_{\odot}$~yr$^{-1}$, while the cooler and less massive ones (e.g. $T_{\rm eff}\lesssim 2\times 10^4$~K and $M \lesssim  7  M_{\odot}$) exhibit weaker winds with  mass-loss rates lower even by three orders of magnitude~\citep{2014A&A...564A..70K}. Mass-loss rates as high as $\sim 10^{-7} \, M_{\odot}$~yr$^{-1}$ have also been theoretically predicted for the more luminous and massive B stars \citep{2000A&A...362..295V}. Typical values for the terminal wind velocity are $V_{\rm Be}^\infty \simeq 500 - 5000$~km s$^{-1}$~\citep[e.g.][]{2000ARA&A..38..613K, 2000A&A...362..295V, 2014A&A...564A..70K}. When the
irradiation from the X-ray emitting NS onto the stellar wind is taken into account, slowing down or even suppression of the wind is predicted \citep{2015A&A...579A.111K}. Such feedback effects are, however, not included in our toy model. 
To account for the range of values in stellar wind properties (found theoretically and observationally), we consider five wind models with different mass loss rates and terminal velocities that are summarized in Table~\ref{tab:Gen_models}. For simplicity, we also adopt a common wind temperature of $2.5 \times 10^4$~K, as this does not  affect our main results.

\begin{table}
	\centering
	\caption{Stellar wind models.}
	\label{tab:Gen_models}
	\begin{tabular}{lcc} 
		\hline
		Model & $\dot{M}_{\rm Be}$ [$10^{-9} M_{\odot}\, {\rm yr}^{-1}$] & $V^{\infty}_{\rm Be}$ [km s$^{-1}$] \\
		\hline
		W1 & 1 & 3000 \\
		W2 & 10 & 3000 \\
		W3 & 100 & 3000\\
		W4 & 1 & 1500 \\
		W5 & 1 & 500 \\
		\hline
	\end{tabular}
\\ Note. -- A common wind temperature of $2.5\times10^4$~K is assumed.
\end{table}

Strong disc winds/outflows have been commonly observed during outbursts in low-mass XRBs \citep[e.g.][]{1997A&A...321..776O,2018Natur.554...69T}. However,
persistent mass loss from the disc has been predicted across different spectral states by radiation-hydrodynamic  simulations \citep{2020MNRAS.492.5271H}.
In the case of magnetized NSs, disc outflows are also expected at sufficiently high mass accretion rates \citep{1973A&A....24..337S, 2016MNRAS.458L..10K} and have been observed in ULXPs \citep[e.g.][]{2018A&A...614A..23K,2019MNRAS.487.4355V,2021arXiv210914683K}.
From a theoretical standpoint, if the accretion is locally super-Eddington, namely the mass accretion rate exceeds the critical rate\footnote{This is defined as $\dot M_{\rm Edd}=L_{\rm Edd}/c^2$.}  $\dot M_{\rm Edd}\simeq 2.3 \times 10^{17}\, M_{\rm NS}/(1.4\, M_{\sun})$~g~s$^{-1}$, part of the dissipated energy in the disc is used to launch outflows from within the so-called spherization radius \citep{1973A&A....24..337S}.
Recently, state of the art simulations have revealed that these outflows can be strong enough to result in optically thick funnel-like structures \citep{2021ApJ...917L..31A}. In addition, apart from the super-Eddington regime, outflows can be formed in low accretion rates from the inner disc radius following its interaction with the NS magnetic field lines \citep{2017ApJ...851L..34P}. Thus, it is clear that outflows can occur in a large variety of systems and conditions for BH and magnetized NS accretors.

In contrast to non-magnetized accreting objects, the accretion discs around magnetized NSs do not extend to the innermost stable orbit, but they are truncated at much larger radii due to the interaction with the NS magnetic field. The magnetospheric radius provides an estimate of the disc  inner radius \citep{1977ApJ...217..578G}:
\begin{equation}
R_{\rm M} = \xi \left(\frac{R_{\rm NS}^{12}B_{\rm NS}^4}{2GM_{\rm NS}\dot{M}^2}\right)^{1/7},
\label{eq:Rm}
\end{equation}
where $G$ is the gravitational constant, $R_{\rm NS}$ is the neutron star radius, $B_{\rm NS}$ is the polar magnetic field strength of the NS, $\dot{M}$ is the accretion rate onto the NS, and $\xi\sim 0.5$ is a model-dependent parameter related to the coupling of the NS magnetic field with the disc \citep{2018A&A...610A..46C}. For typical magnetic field strengths in X-ray pulsars (e.g., $10^{12}$ G), very high mass accretion rates are required 
(e.g., $>10 \, \dot{M}_{\rm Edd}$ or $L_{\rm X}\gtrsim 10^{39}$~erg s$^{-1}$) to push  the NS magnetosphere within the spherization radius.  

In our toy model, we do not specify the origin of the NS outflow and postulate that  it is present regardless of the accretion regime. The outflow is also assumed to be spherically symmetric and homogeneous (i.e. non clumpy).  The rate at which mass is carried away from the NS is parameterized as $\dot{M}_{\rm NS}=\eta_{\rm out} L_{\rm X}/(\epsilon c^2)$, where  and $c$ is the speed of light, $\epsilon\sim 0.2$ is the accretion efficiency onto the NS \citep[e.g.,][]{2019MNRAS.488.5225V}, and $\eta_{\rm out}<1$ is the fraction of the accreted mass lost through the outflow. The NS outflow speed is considered constant and proportional to the Keplerian velocity at $R_{\rm M}$, which is a proxy of the inner disc radius, namely $V_{\rm NS}^\infty=\chi \sqrt{G M_{\rm NS}/R_{\rm M}}$. Here, $\chi\sim1-3$ to allow for an accelerating outflow.

\begin{figure}
    \centering
    \includegraphics[width=0.49\textwidth]{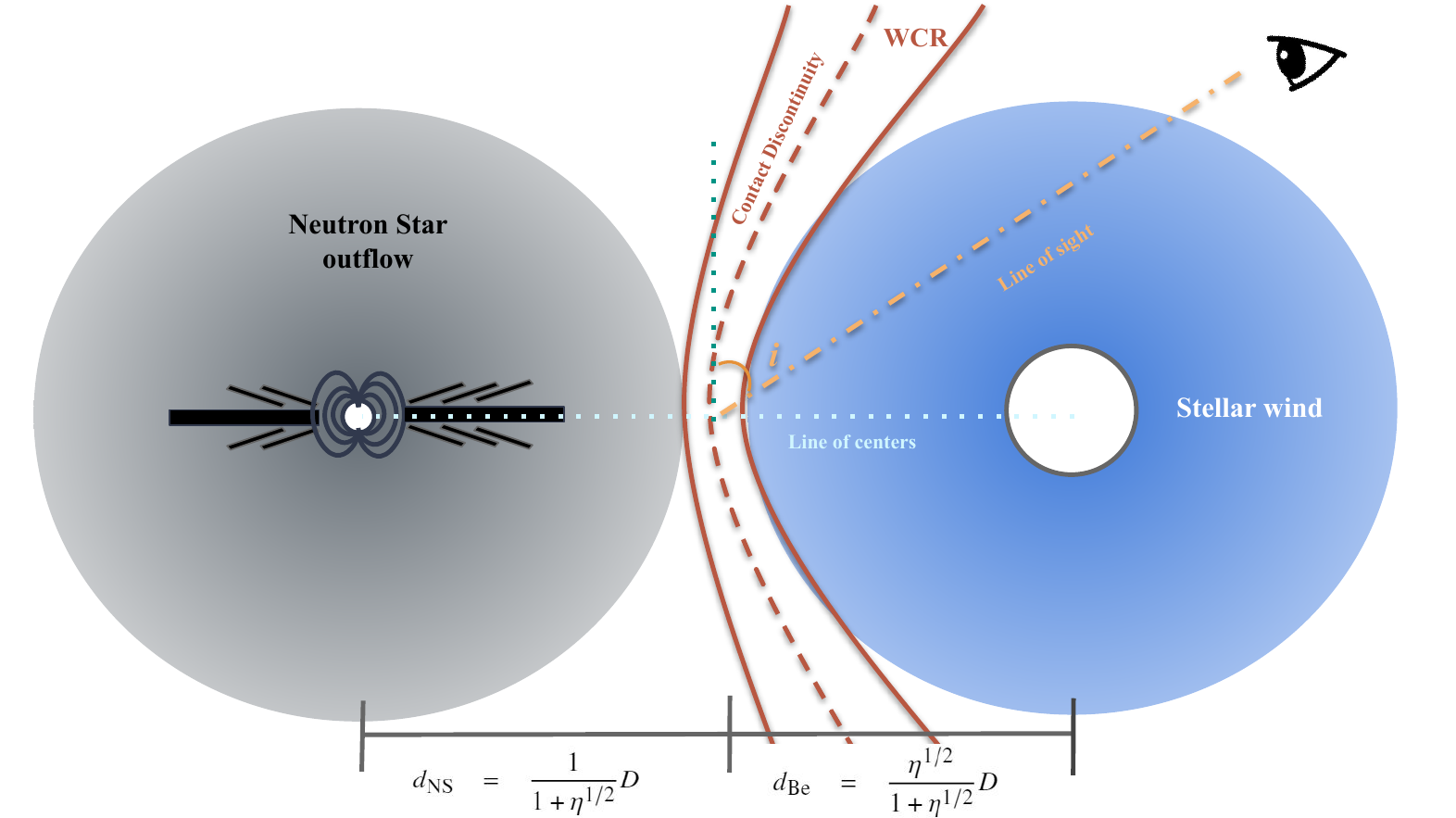}
    \caption{A schematic illustration of our toy model (not in scale).}
    \label{fig:sketch}
\end{figure}

If the binary separation distance is large enough, the two stellar winds can collide at high speeds, forming the so-called wind colliding region (WCR) \citep{1992ApJ...389..635U}. Neglecting the effects of the orbital motion and under the assumption of spherically symmetric and homogeneous outflows, the WCR is almost spherical at small angles as measured from the line connecting the centers of the two stars, and becomes conical only at large angles \citep{1992ApJ...389..635U, 1993ApJ...402..271E}.
The apex of the shock  on the axis that connects the two stars is located at a distance $d_{\rm NS}$ and $d_{\rm Be}$ from the NS and Be star, respectively (see also Fig.~\ref{fig:sketch}). Assuming spherical colliding winds, these distances are written as \citep{1993ApJ...402..271E}, 
\begin{eqnarray}
d_{\rm NS}&=& \frac{1}{1+\eta^{1/2}}D  \\ 
d_{\rm Be}&=& \frac{\eta^{1/2}}{1+\eta^{1/2}}D
\end{eqnarray}
where $D$ is the binary separation distance and $\eta$ is the wind momentum ratio that is defined as
\begin{equation}
\eta=\frac{\dot{M}_{\rm Be}V^{\infty}_{\rm Be}}{\dot{M}_{\rm NS}V^{\infty}_{\rm NS}}.
\label{eq:eta}
\end{equation}
If $\eta >  1$ (i.e., the Be wind is stronger than the NS outflow), the shock will be wrapped around the NS and $d_{\rm NS} < d_{\rm Be}$. However, for typical Be stellar wind parameters and super-Eddington accretion rates, the opposite situation may be realized, as illustrated in Fig.~\ref{fig:sketch}. In both cases, the shock radius, $R_{\rm sh}$, is measured from the center of the star that is closer to the shock, i.e., $R_{\rm sh}=\min(d_{\rm NS}, d_{\rm Be})$. Parameter values leading to $R_{\rm sh} < R_{\rm NS}$ or $R_{\rm sh} \lesssim 20 R_{\odot}$ are ignored as being non physical. The latter is a loose limit for the radial distance in which the wind acceleration is completed~\citep[e.g.][]{1986ApJ...311..701F, 1999isw..book.....L}.

\subsection{Radio emission from the system}\label{sec:radiation}
Astrophysical shocks can accelerate charged particles to high energies as indicated by the associated non-thermal emission. First-order Fermi acceleration, according to which particles gain energy through repeated shock crossings mediated by scattering off magnetic perturbations on both sides, is thought to be the main acceleration mechanism at play \citep{1978ApJ...221L..29B, 1978MNRAS.182..147B,1983RPPh...46..973D}. The energy distribution of accelerated particles is found to be a power law, i.e. $N(E)\propto E^{-p}$, with the index $p$ depending on the shock compression ratio $r$. For strong shocks, $r = 4$ and $p\simeq 2$. 

Strong shocks formed by the colliding stellar winds in massive binary systems have been discussed as promising sites of particle acceleration and non-thermal radiation \citep[e.g.][]{1993ApJ...402..271E, 2003A&A...409..217D, 2021MNRAS.504.4204P}.
In our toy model, we assume that relativistic electrons are produced by first-order Fermi acceleration at the shocks of the colliding stellar wind and NS outflow, and compute the synchrotron emission produced in the region of the shocked stellar wind\footnote{For the calculation of non-thermal emission we consider only the spherical part of the WCR.}.

A vital quantity for this calculation is the energy density of the shocked wind region $U_{\rm Be}$. In order to calculate said quantity we employ our knowledge that for a strong shock
we have $U_{\rm Be}=(9/8)\rho (V^{\infty}_{\rm Be})^2$, where $\rho=\dot{M}_{\rm Be}/4\pi d^2_{\rm Be} V_{\rm Be}^\infty$ the pre-shock wind density at distance $d_{\rm Be}$. Thus, in terms of $\eta$ and our wind parameters we have 
\begin{equation}
U_{\rm Be}=\frac{9}{32\pi}\dot{M}_{\rm Be}V^{\infty}_{\rm Be}D^{-2}(1+\eta^{-1/2})^{2}.
\label{eq:UBe}
\end{equation}
Accordingly, we can calculate the total electron energy density of the shocked electron population by attributing to it a fraction of the total energy density, namely $U_{\rm e}=\epsilon_{\rm e} U_{\rm Be}$, while assigning the rest to the energy density of the post-shock magnetic field, that is $U_{\rm B}=\epsilon_{\rm B} U_{\rm Be}$.

The energy distribution of the accelerated electrons per unit volume is modelled as a power law extending from a minimum to a maximum Lorentz factor ($\gamma_{\min}$ and $\gamma_{\max}$, respectively) with a slope of $p$, namely
$n(\gamma)\equiv {\rm d}N/{\rm d}\gamma {\rm d}V=K_{\rm e}\gamma^{-p}$, $\gamma_{\min}\le \gamma \le \gamma_{\max}$. Here, $K_{\rm e}$ is the normalization factor that is directly related to the total electron energy density via $U_{\rm  e}=m_e c^2 \int_{\gamma_{\min}}^{\gamma_{\max}} n(\gamma)  (\gamma-1) {\rm d}\gamma$. Even though index $p$ is related to the shock compression ratio, our model is not designed to describe the acceleration process. Hence, we consider $p\ge 2$ a free parameter of the model. For the same reasons, we also let $\gamma_{\min}$ to be a free parameter and assume that $\gamma_{\max} \gg \gamma_{\min}$.

Relativistic electrons in the WCR can cool (i.e. lose energy) due to synchrotron radiation and inverse Compton  (IC) scattering on stellar photons. As a result of synchrotron and/or IC cooling (in the Thomson regime), the power-law slope of the accelerated electron population steepens by one, i.e. $N(\gamma)\sim\gamma^{-p-1}$, for values of $\gamma$ above a critical Lorentz factor, known as the cooling Lorentz factor. This is estimated by balancing the radiative loss timescale, $t_{\rm loss}$, with a typical timescale, $t_{\rm dyn}$, on which particles are advected along the contact discontinuity. The cooling timescale for an electron with Lorentz factor $\gamma$ due to synchrotron and IC losses in the Thomson regime\footnote{Electrons with Lorentz factors up to $\sim m_{\rm e} c^2/\epsilon \simeq 5 \times 10^5 (1~{\rm eV}/\epsilon)$ can up-scatter stellar photons of energy $\epsilon$ in the Thomson regime.} is $t_{\rm loss}\approx 3 \pi m_{\rm e} c / (4 \sigma_{\rm T} (U_{\rm B} + U_{\rm ph}) \gamma)$. Here, $U_{\rm ph} \approx L_{\rm Be}/ 4 \pi c \, d_{\rm Be}^2$ is a proxy for the energy density of stellar photons that are up-scattered by the electrons in the WCR~\citep[e.g.][]{2006A&A...446.1001P}. We adopt $L_{\rm Be} = 10^4 L_{\odot}$ as a typical value for the bolometric luminosity of the Be star. For simplicity, we set $ t_{\rm dyn}\approx R_{\rm sh}/(r V)$ where $V \approx V^{\infty}_{\rm Be}/r$ is the post-shock flow velocity and $r\approx 4$ is the shock compression ratio. In reality, the advection velocity will depend on the position along the contact discontinuity \citep[e.g.][]{2015A&A...577A.122J}, but this cannot be captured by the single zone nature of our toy model. The cooling Lorentz factor due to synchrotron and IC losses in the Thomson limit is then written as
\begin{equation}
\gamma_{\rm c}=\frac{3 m_{\rm e}c V^{\infty}_{\rm Be}}{4\sigma_{\rm T}(U_{\rm B}+U_{\rm ph}) R_{\rm sh}},
\label{eq:gcool}
\end{equation}
where $m_{\rm e}$ is the electron mass. For most parameter combinations we consider, we find that $\gamma_{\rm c} > \gamma_{\min}$ \citep[slow cooling regime,][]{1998ApJ...497L..17S} with lower $\gamma_{\rm c}$ values obtained whenever the WCR is pushed very close to the Be star (i.e. for strong NS outflows and/or weak Be winds). In the latter case, IC cooling dominates the energy losses and can even lead to $\gamma_{\rm c}\ll \gamma_{\min}$ \citep[fast cooling regime,][]{1998ApJ...497L..17S}. Fast-cooling electrons would be confined to regions close to the shocks and would not occupy the full width of the WCR, resulting in a smaller emitting volume than the one for slow-cooling electrons~\citep[e.g.][]{2006A&A...446.1001P}. This effect is not captured by our toy model. Moreover, the point-source approximation for the calculation of $U_{\rm ph}$ would break down when the WCR would form close to the Be star. Notably, most of the fast cooling cases are realized for $d_{\rm Be} < 20 R_\odot$ (non-physical solutions). For these reasons, we focus on the slow-cooling regime from this point on.

Having specified the above, we are able to derive the synchrotron emission accounting for possible suppression stemming from attenuation processes, such as synchrotron self-absorption within the shocked wind region and free-free absorption from the unshocked stellar wind. Moreover, in the presence of a background plasma, the synchrotron emission of a relativistic particle with Lorentz factor $\gamma$ will be suppressed at low enough frequencies, since the beaming of the radiation is not as strong at frequencies $\gamma \omega_{\rm p}$, where $\omega_{\rm p}$ is the plasma frequency; this is known as the Razin effect \citep{razin60}.

\begin{table*}
	\centering
	\caption{Default values of model parameters.}
	\label{tab:f_par}
	\begin{tabular}{lccr} 
		\hline
		Parameter & Symbol & Value\\
		\hline
        NS magnetic field strength & $B_{\rm NS}$ & $10^{12}$~G \\ 
        NS radius & $R_{\rm NS}$ & $10^6$~cm \\ 
        NS mass & $M_{\rm NS}$ & $1.4\, M_\odot$ \\ 
        Be bolometric luminosity & $L_{\rm Be}$ & $10^4 L_{\odot}$ \\
        Ratio of NS outflow velocity and Keplerian velocity at inner radius & $\chi$ & 2\\
        Fraction of accreted mass lost through outflows & $\eta_{\rm out}$ & 0.01\\
        Radiative efficiency of accretion flow & $\epsilon$ & 0.2\\
        Fraction of magnetic field energy density & $\epsilon_{\rm B}$ & 0.5 \\
        Fraction of relativistic electron energy density & $\epsilon_{\rm e}$ & 0.5 \\
        Power-law index of electron energy distribution & $p$ & 2 \\
        Minimum electron Lorentz factor & $\gamma_{\min}$ & 10 \\
        Binary separation distance & $D$ & $10^3 R_\odot$ \\
        Inclination angle & $i$ & $\pi/4$\\
		\hline
	\end{tabular}
\end{table*}

\begin{figure*}
    \includegraphics[width=0.49\textwidth]{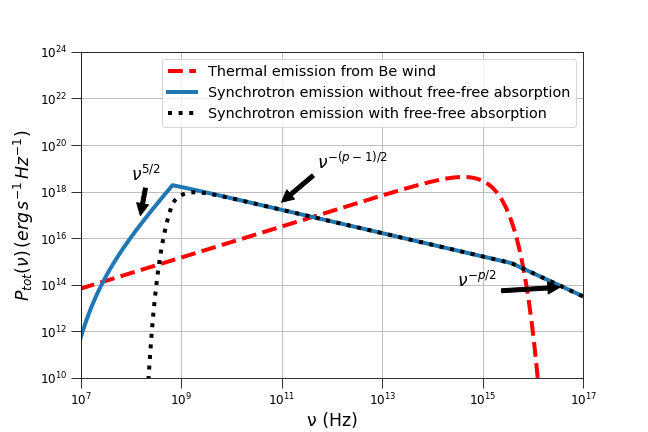}
    \includegraphics[width=0.49\textwidth]{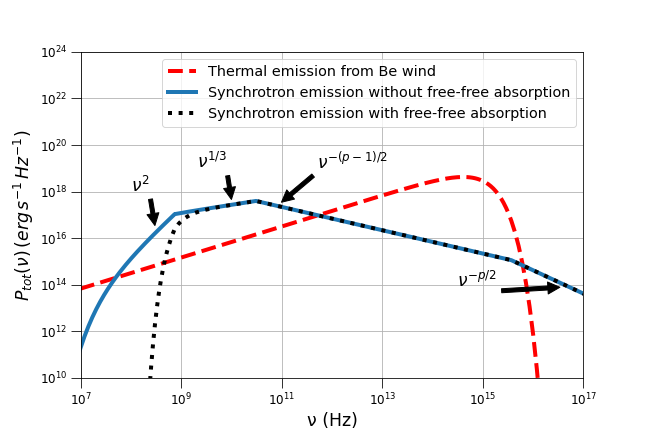}
    \caption{Spectral decomposition of the thermal and non-thermal emission of the system for $\gamma_{\min}=10$ (left panel) and $\gamma_{\min}=100$ (right panel). In both cases the effect of the free-free absorption on the synchrotron emission is noted. Here, the stellar wind model W3 was used and $L_{\rm X}=10^{38}$~erg s$^{-1}$ was assumed. All other parameters used are listed in Table~\ref{tab:f_par}.}
    \label{fig:SectraDecomp}
\end{figure*}

We can then write the intensity of the synchrotron radiation at frequency $\nu$  as
\begin{equation}
I_{\rm \nu}^{\rm syn}= S_{\rm \nu}^{\rm syn}(1-e^{-\tau_{\rm \nu}^{\rm  ssa}})e^{-\tau_{\rm \nu}^{\rm ff}}e^{-\nu_{\rm R}/\nu}
\end{equation}
where $I_{\rm \nu}^{\rm syn}$ is in units of $\rm erg\, cm^{-2} s^{-1}sr^{-1}Hz^{-1}$.
Here, we have the source function $S_{\rm \nu}^{\rm syn}=j_{\rm \nu}^{\rm syn}/a_{\rm \nu}^{\rm ssa} $ with $j_{\rm \nu}^{\rm syn}$ being the synchrotron emissivity, $a_{\rm \nu}^{\rm ssa}$ the synchrotron self-absorption coefficient, $\tau_{\rm \nu}^{\rm ssa}=a_{\rm \nu}^{\rm ssa} R_{\rm sh}/4$ the optical depth for synchrotron self-absorption, $\tau_{\rm \nu}^{\rm ff}$ the optical depth for free-free absorption, and $\nu_{\rm R}\simeq 20 n_{\rm sh}/B$ a critical frequency below which the Razin effects becomes important \citep[e.g.][]{2003A&A...409..217D, 2016MNRAS.460...44P}. Here, $n_{\rm sh} = 4 \rho/(\mu m_{\rm p})$ represents the post-shock electron number density and $\mu$ is the mean molecular weight (0.6 for solar composition).  The free-free optical depth along a path starting from the apex of the shock (see Fig.~\ref{fig:sketch}) can be written as~\citep{1975MNRAS.170...41W}
\begin{equation}
\tau_{\rm \nu}^{\rm ff}= K(\nu,T)A^2 \left(d_{\rm Be} \rm \cos i\right)^{-3}\int_{- i}^{\pi/2}\cos^2\theta \, {\rm d}\theta,
\end{equation}
with 
\begin{equation}
    A= \frac{\dot{M}_{\rm Be}}{4\pi\mu m_{\rm H}V^{\infty}_{\rm Be} } 
    \label{eq:A}
\end{equation}
and 
\begin{equation}
K(\nu,T)=3.7 \times 10^8 \, (1-e^{ -h\nu/kT})Z^2g(\nu,T)T^{-1/2}\nu^{-3} 
\label{eq:Kv}
\end{equation}
in cgs units.
In the above expressions $i$ is the inclination angle  (see also Fig.~\ref{fig:sketch}), $m_{\rm H}$ the mass of the hydrogen atom, $Z$ the ion charge, $g(\nu,T)$ is the Gaunt factor, and $T$ is the temperature of the Be wind. A more general expression of $\tau^{\rm ff}_\nu$ can be found in Appendix~\ref{app:tff}.

Knowing the above, we can calculate the luminosity per unit frequency as
\begin{equation}
L_{\rm \nu}^{\rm syn}=4\pi R_{\rm sh}^2 I_{\rm \nu}^{\rm syn} f_{\rm V} 
\end{equation}
where $f_{\rm V}$ a fraction of a whole sphere. In our case, we take this to be $f_{\rm V}=0.25$, corresponding to a spherical wedge from $-\pi/4$ to $+\pi/4$ above and below the line of centers (see Fig.~\ref{fig:sketch}). For larger angles the approximation for a head-on collision of the winds at each point begins to break down \citep[e.g.][]{2021MNRAS.504.4204P}. Moreover, we calculate the thermal Brehmsstrahlung of the Be wind as \citep{1975MNRAS.170...41W}
\begin{equation}
L_{\rm \nu}^{\rm th}=141.9 \, \zeta^{2/3} A^{4/3}B(\nu,T)K^{2/3}(\nu,T),
\end{equation}
where the electron number density is assumed to be equal to $\zeta$ times the ion number density, and $B(\nu,T)$ the Planck function. Finally, the total observed luminosity per unit frequency is given  by 
\begin{equation}
    L_{\rm \nu}^{\rm obs}=L_{\rm \nu}^{\rm th}+L_{\rm \nu}^{\rm syn}.
    \label{eq:Lnu}
\end{equation}

\begin{figure*}
    \includegraphics[width=0.49\textwidth]{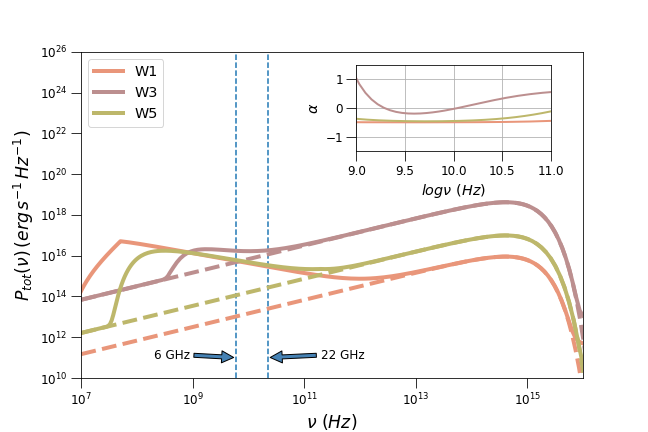}
    \includegraphics[width=0.49\textwidth]{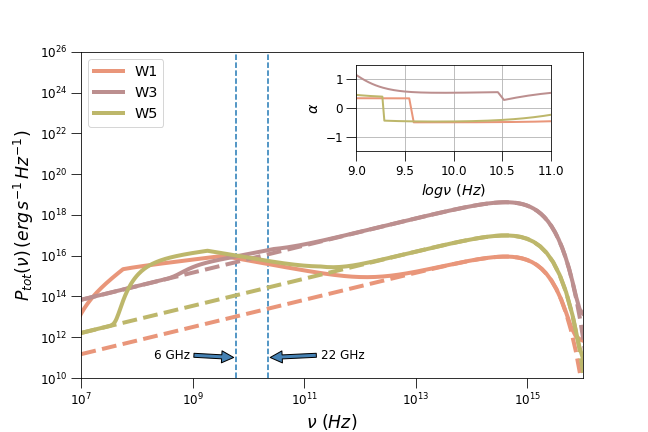}
    \caption{Radio-to-UV spectrum for three wind models and $L_{X}=10^{37}$~erg s$^{-1}$. Left and right panels show results for $\gamma_{\min}=10$ and 100, respectively. Thick solid and dashed lines show the total emission and thermal wind emission, respectively. Vertical dashed lines indicate two characteristic radio frequencies. In a strong wind scenario, like W3, the 6~GHz emission is dominated by the Be stellar wind. Inset plots show $\alpha \equiv d\log P_{\rm tot}(\nu)/d\log\nu $ as a function of frequency between 1 and 100 GHz.}
    \label{fig:RadioSpec37}
\end{figure*}

\begin{figure*}
    \includegraphics[width=0.49\textwidth]{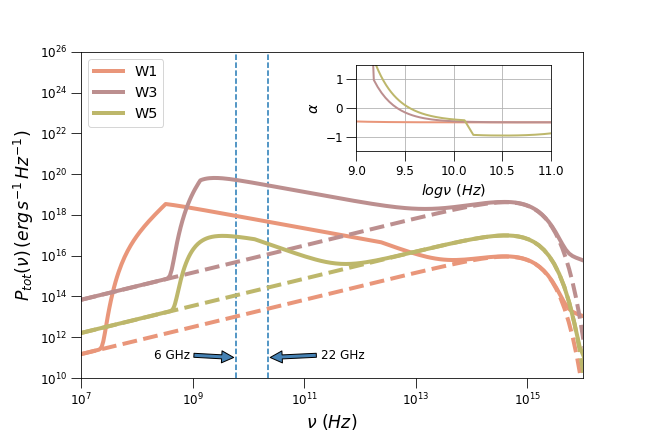}
    \includegraphics[width=0.49\textwidth]{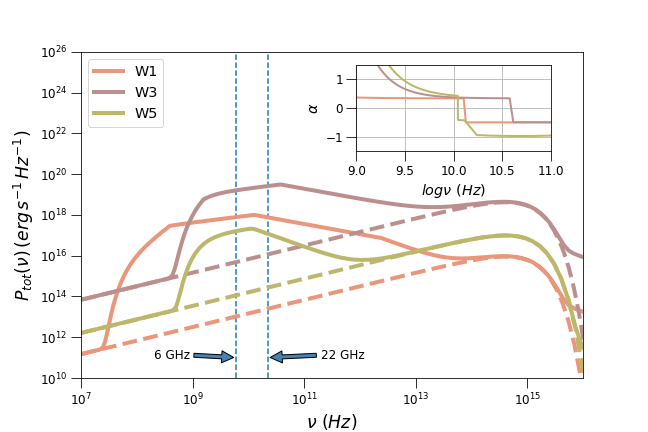}
    \caption{Same as in Fig.~\ref{fig:RadioSpec37} but for  $L_{\rm X}=10^{39}$~erg s$^{-1}$.}
    \label{fig:RadioSpec39}
\end{figure*}

\section{Results}\label{sec:results}
Having described the setup of our toy model we move next in the calculation of the radio emission of the system for the wind models listed in Table~\ref{tab:Gen_models}. The parameter values used in all figures, if not otherwise noted, are summarized in Table~\ref{tab:f_par}. 

\subsection{Spectra}
Spectra of the total emission computed using equation~(\ref{eq:Lnu}) are displayed for two values of the minimum electron Lorentz factor, i.e. $\gamma_{\rm min}$=10 and 100 on the left and right panels of Fig.~\ref{fig:SectraDecomp}, respectively. The parameters describing the stellar wind are those of model W3, while $L_{X}=10^{38}$~erg s$^{-1}$ is adopted. For these parameter values, the shock is wrapped around the Be star, as shown in Fig.~\ref{fig:sketch}. The total spectrum is composed of two components, namely the non-thermal synchrotron emission (dotted line) and the free-free emission of the Be stellar wind (dashed line). The synchrotron spectrum without accounting for free-free absorption is also shown for comparison (solid line). The spectral curvature seen below $\nu_{\rm ssa}\sim 1$~GHz is due to the Razin suppression. For most parameters considered in this work, free-free absorption is more important than the Razin suppression.
The ordering of the synchrotron self-absorption frequency and the minimum characteristic synchrotron frequency ($\propto B\gamma^2_{\min}$) changes for the two values of $\gamma_{\min}$ considered here. This is illustrated by the change in slope of the synchrotron self-absorbed part of the spectrum.
For the adopted parameter values, the non-thermal emission from the shocked wind starts to dominate the thermal wind emission at $\nu \sim 1$~GHz. The transition frequency from the thermal to the non-thermal component strongly depends on the parameters, as we will illustrate later in this section. At lower frequencies the free-free opacity becomes much larger than unity, thus significantly reducing the synchrotron flux. The thermal emission from the ionized stellar wind dominates again the overall emission between the infrared and UV wavelengths, with a peak at $\sim 10^{15}$~Hz ($\sim 4$~eV), which is determined by the wind temperature.

The effects of the stellar wind parameters (i.e. mass loss and terminal velocity) on the combined radio spectra of the system are illustrated in Figs.~\ref{fig:RadioSpec37} and \ref{fig:RadioSpec39} for $L_{X}=10^{37}$~erg s$^{-1}$ and $10^{39}$~erg~s$^{-1}$, respectively. Two characteristic observing radio frequencies are indicated with vertical dashed lines. 
Within the figures the spectral indices for radio frequencies are also noted. 
The competition between the thermal and non-thermal emission processes determines the observed radio spectrum. For instance, for a strong Be wind scenario as in model W3, and sufficiently low accretion rates, the radio luminosity at 6 and 22 GHz is dominated by the thermal wind emission (compare purple lines on the right-hand side panels in Figs.~\ref{fig:RadioSpec37} and \ref{fig:RadioSpec39}). 

\begin{figure}
    \centering
    \includegraphics[width=0.49\textwidth]{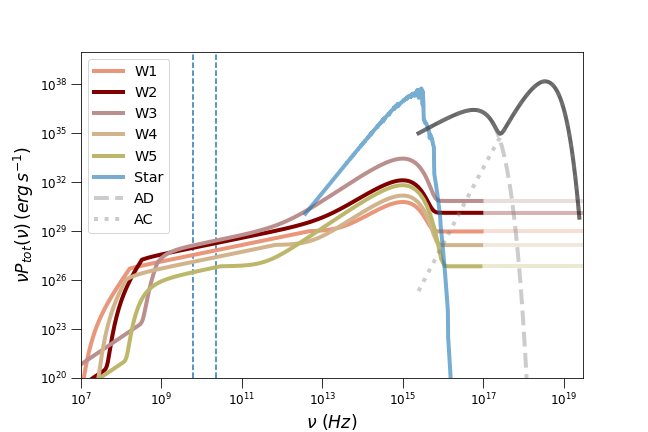}
    \caption{Composite radio-to-X-ray spectra for a fiducial HMXB with $L_X=L_{\rm Edd}=1.8\times10^{38}$~erg s$^{-1}$. The spectrum from the stellar surface of the massive companion is overplotted with solid blue line. The X-ray spectra (without absorption) from the accretion column (AC, dotted grey line) and the accretion disc (AD, dashed grey line) are also shown for comparison purposes. Vertical dashed lines mark the 6~GHz and 22~GHz frequencies. }
    \label{fig:composite_spec}
\end{figure}

Before closing this section, it would be useful to put the spectral predictions of our model into perspective. We therefore compare the thermal and non-thermal emission of our model for the default parameter values (Table~\ref{tab:f_par}) and the various wind models presented in Table~\ref{tab:Gen_models} with the spectrum of a massive companion star and the X-ray emission powered by accretion. As a representative example, we used the Kurucz spectral model for a massive star with  $T_{\rm eff}=2.5\times10^4$~K, radius $5 \, R_{\odot}$, gravitational acceleration $\log g=5$ (in cgs units), turbulent velocity $v_{\rm turb}=2$~km s$^{-1}$, and solar composition\footnote{\url{https://wwwuser.oats.inaf.it/castelli/grids.html}}. X-rays are produced both in the accretion disc  and the accretion column above the NS surface. As an illustrative example, we considered the \citet{2007ApJ...654..435B} model that computes the spectrum of the accretion column. The model assumes a cylindrically collimated radiation-dominated radiative shock in the accretion flow confined by the NS magnetic field and has been applied to prototypical X-ray pulsar Her X-1 \citep[][]{2016ApJ...831..194W}. 
For the accretion column spectrum we used $B=10^{12}$~G and $\dot{M}=10^{18}$~g s$^{-1}$, while the rest physical parameters where fixed to their typical values \citep{2016ApJ...831..194W}.
The spectrum of the accretion disc is approximated by a multi-temperature black body with inner disc radius $R_{\rm in}= R_{\rm M} \simeq 500$~km and temperature $T_{\rm in}=0.1$ keV. 
Our results are presented in Fig.~\ref{fig:composite_spec}. Stellar radiation dominates the emission from near-infrared\footnote{We neglected possible contribution of the decretion disc in this energy range.} (NIR) to optical/UV frequencies. This is a major difference with low-mass X-ray binaries where the NIR/optical emission has contributions from the disc and the jet in low-hard states~\citep[e.g.][]{2003A&A...397..645M, 2013MNRAS.429..815R}. The synchrotron spectrum of relativistic electrons in the WCR can extend to X-rays (transparent solid coloured lines) for sufficiently high $\gamma_{\max}$ values.  Here, $\gamma_{\max}$ is computed by balancing the synchrotron loss rate with the acceleration rate (in the Bohm limit). Hence, the spectrum extends to $\sim 160$~MeV, known as the synchrotron burn-off limit \citep{1996ApJ...457..253D}. IC cooling is not included in the determination of $\gamma_{\max}$ because scatterings off stellar photons by electrons with such high Lorentz factors happen deep in the Klein-Nishina regime. Even though synchrotron emission is guaranteed in the X-ray regime, its luminosity is orders of magnitude smaller than the X-ray luminosity from the accretion column and the inner regions of the accretion disc.

\begin{figure*}
    \includegraphics[width=0.49\textwidth]{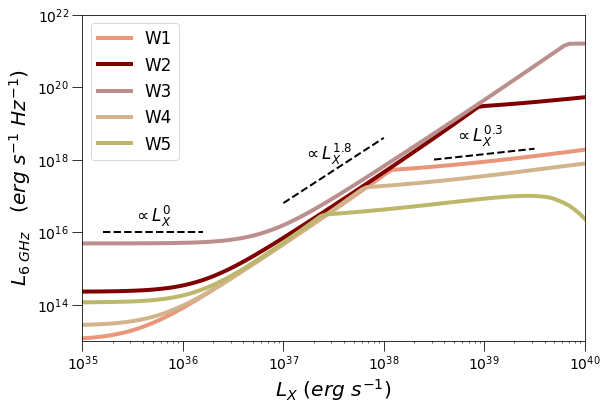}
    \includegraphics[width=0.49\textwidth]{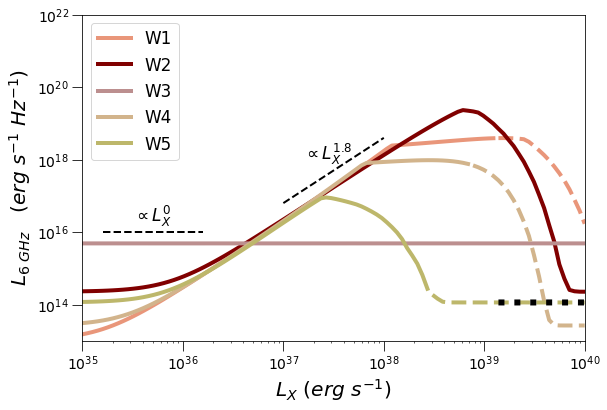}
    \caption{Luminosity at 6~GHz plotted against the X-ray luminosity due to accretion for the wind models presented in Table~\ref{tab:Gen_models}. Left and right panels show results for $D=1000~R_{\odot}$ and $100~R_{\odot}$, respectively. Dashed lines represent nonphysical solutions with $R_{\rm sh} < 20 R_{\odot}$. Dotted black lines correspond to the fast electron cooling regime where our analysis is not valid. For strong stellar winds, like in W3, the emission can be dominated by the wind thermal radiation, which is independent of the NS X-ray luminosity (see right panel).}
    \label{fig:LCex}
\end{figure*}

\subsection{The $L_R - L_X$ diagram}
Examining our model predictions for the emission at 6~GHz for a wide range of X-ray luminosities, we can construct the $L_{R}$ versus $L_{X}$ plot, as shown in Fig.~\ref{fig:LCex} for two values of the binary separation distance. In all cases considered, there is a range of X-ray luminosities where the free-free emission from the Be wind dominates, hence $L_{R}\propto L_X^0$. For higher X-ray luminosities, a power-law relation is established, i.e. $L_{R}\sim L_{X}^{b}$ with $b \simeq 1.7-1.8$. In this regime, the 6 GHz luminosity is attributed to the optically thin part of the (uncooled) synchrotron spectrum (see also left-hand side panels in Figs.~\ref{fig:RadioSpec37} and \ref{fig:RadioSpec39}). For even higher values of $L_X$ we still find a positive correlation between radio and X-ray luminosities, but with $b \sim 0.3$, as shown in the left-hand side panel of Fig.~\ref{fig:LCex}. 

The change in slope $b$ signals a change in the wrapping orientation of the WCR, i.e. the shock is wrapped around the NS for low enough $L_X$ ($\eta > 1$), but then is pushed toward the Be star as the accretion rates become higher ($\eta < 1$). The value of $L_X$ that corresponds to this transition is found by setting $\eta = 1$ (see equation \ref{eq:eta})
\begin{eqnarray}
L_{X, \rm br} &\simeq& 4\times10^{37} \ \epsilon_{-1} \ \xi^{7/16}_{-0.3} \  \mu_{\rm NS, 30}^{1/4} M_{\rm NS, 0.15}^{-1/2} \nonumber \\ & & \left(\frac{\dot{M}_{\rm Be, -9} V_{\rm Be, 3}^{\infty}}{\eta_{\rm out,-1} \chi_0}\right)^{7/8} \rm erg~s^{-1}
\label{eq:Lxcr}
\end{eqnarray}
where $\mu_{\rm NS}\equiv B_{\rm NS} R_{\rm NS}^3$ is the NS dipole magnetic moment. We also introduced the notation $q_x = q/10^x$, with the mass loss rate in units of $M_\odot$ yr$^{-1}$,  the wind velocity in units of km~s$^{-1}$, and the NS mass in units of the solar mass. Note that for typical parameter values, the transition occurs close to the Eddington luminosity of a NS.

As long as the radio frequency of interest falls in the optically thin (and uncooled) part of the synchrotron spectrum, the relation $L_R \propto L_X^b$ can be derived analytically, as shown in Appendix~\ref{app:LrLx}, 
\begin{eqnarray}
L_R \propto \left \{ 
\begin{tabular}{c c}
$L_X^{12/7}$,     & $\eta \gg 1$ or $L_{X} \ll L_{X, \rm br}$ \\  \\
$L_X^{2(p-1)/7}$, & $\eta  \ll 1$ or $L_{X} \gg L_{X, \rm br}$
\end{tabular}
\right.
\label{eq:LrLx}
\end{eqnarray}
For the dependencies of $L_R$ on the model parameters, we refer the reader to Appendix~\ref{app:LrLx}.

Similar results are found for smaller binary separation distances, as shown in the right panel of Fig.~\ref{fig:LCex}. However, as the system becomes more compact, the shallow part of the radio X-ray luminosity curve (with $b\sim 0.3$) becomes truncated due to the increasing role of absorption. In other words, the abrupt steepening indicates that the radio emission at 6 GHz becomes optically thick. Moreover, for high enough $L_X$ values (depending on the wind model) the WCR is pushed very close to the Be star, making these solutions not physically meaningful (dashed coloured lines). Note also that for weak Be winds (like in model W5) and high X-ray luminosities (i.e. strong NS outflows), GHz-emitting electrons can be fast cooling due to strong IC losses (black dotted lines).

\subsection{Effects of model parameters} 

\begin{figure*}
\includegraphics[width=\textwidth]{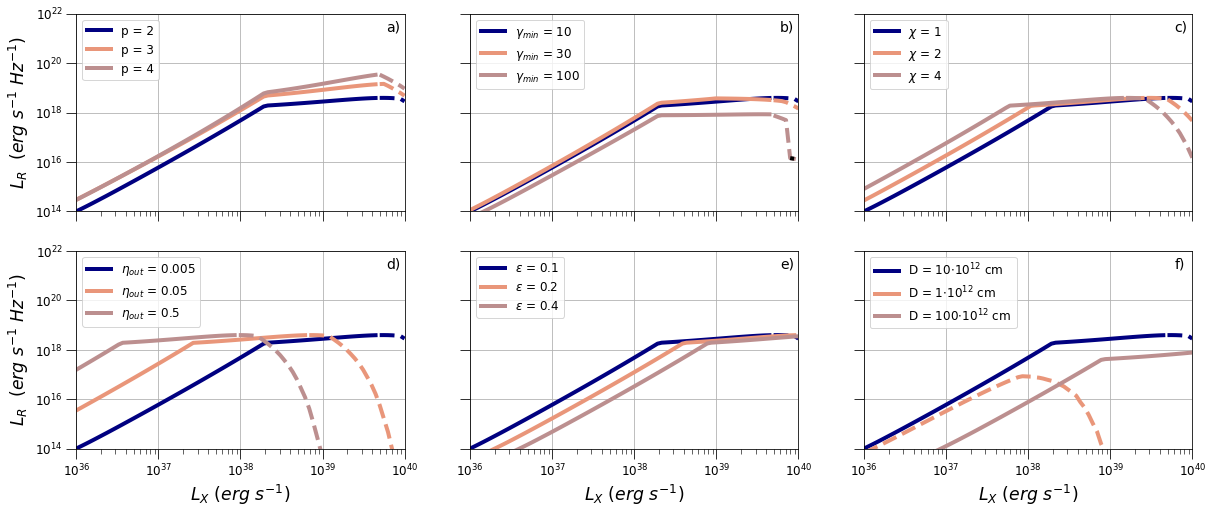}
    \caption{Impact of the model parameters on the $L_R-L_X$ diagram for the W1 model: a) power-law slope electron distribution, b) minimum Lorentz frequency of the electron distribution, c) deviation of the NS outflow velocity from the Keplerian value at the inner disc radius, d) fraction of the NS accretion rate that gets lost through outflow, e) mass-to-energy conversion efficiency of the accreting NS, f) the binary distance. In all panels, dashed lines represent non physical solutions with $R_{\rm sh} \le 20 R_{\odot}$. Dotted black lines correspond to the fast electron cooling regime where our analysis is not valid.}
    \label{fig:multiple}
\end{figure*}

The impact of the main model parameters on the $L_R-L_X$ diagram is illustrated in Fig.~\ref{fig:multiple}. In each panel, we vary one model parameter as indicated in the inset legends, while keeping the others fixed to the following values: $p=2$, $\gamma_{\min}=3$, $\chi=1$, $\eta_{\rm out}=0.005$, $\epsilon=0.1$, and $D=10^{13}$~cm. Here, the wind model W1 is used.  

Examining the different panels of Fig.~\ref{fig:multiple} we can make the following remarks. First, a steeper power-law distribution in electrons (panel a) implies a larger number of accelerated electrons at lower energies. Hence, the number of particles radiating at radio frequencies is larger for $p=4$ than for $p=2$, when all other parameters are fixed, and an increase in $L_{R}$ is found. Notice also the change in slope $b$ of the power-law segment above the critical X-ray luminosity (see also equations~\ref{eq:Lxcr} and \ref{eq:LrLx}). The synchrotron luminosity at 6 GHz depends also on the chosen value of $\gamma_{\min}$ (panel b). For higher values of $\gamma_{\min}$, the optically thin synchrotron emission at a fixed frequency (here, 6 GHz) decreases due to the smaller number of low-energy electrons radiating at that frequency. For the same physical conditions, the electron cooling Lorentz factor (see equation \ref{eq:gcool}) can be smaller than $\gamma_{\min}$, if the latter is sufficiently large (dotted black lines for $\gamma_{\min}=100$). The acceleration of the NS outflow to a speed $\chi$ times larger than the Keplerian velocity at the inner disc radius does not affect much the shape of the $L_R-L_X$ curve (panel c). Instead, an increase of $\chi$ directly implies a decrease in $\eta$ (equation~\ref{eq:eta}). Therefore, the whole curve shifts to the left with increasing $\chi$. Similarly, the curves shift to the left and to the right with increasing $\eta_{\rm out}$ and decreasing $\epsilon$, respectively (panels d and e). The binary separation distance has the biggest effect on the curves, as shown in panel f. For compact enough binaries, free-free attenuation by the stellar wind becomes very strong, thus reducing the 6~GHz radio luminosity. At the same time, the WCR can form very close to Be star, where the assumption for wind acceleration to terminal velocity breaks down.

Inspection of Fig.~\ref{fig:multiple} shows that the radio luminosity computed at $L_{X, \rm br}$ can be used as a proxy of the maximum expected value from an XRB (but see also panel a). This is independent of the parameters describing the NS outflow, namely $\chi, \eta_{\rm out}$ and $\epsilon$, as shown in panels c to e. Indeed, using equations (\ref{eq:Lxcr}) and (\ref{eq:LrLx-1}) we find
\begin{eqnarray}
\nu_R L_R \large |_{L_{X, \rm br}} & \simeq & 2 \times 10^{27} \nu_{R,10}^{\frac{1}{2}} \left(\dot{M}_{\rm Be,-9} V_{\rm Be, 3}^{\infty} \right)^{\frac{7}{4}} \epsilon_{\rm B, -0.3 }^{\frac{3}{4}}  \nonumber \\
& & \epsilon_{\rm e, -0.3}  D_{13}^{-\frac{1}{2}} \ \ln^{-1}\left(\frac{\gamma_{\max}}{\gamma_{\min}}\right) {\rm erg \, s^{-1}}
\end{eqnarray}
where we assumed $p=2$ for simplicity. Thus, the maximum radio luminosity at GHz frequencies depends mostly on the stellar wind momentum and the binary separation distance, and can range between $10^{27}-10^{28}$~erg s$^{-1}$ for typical parameter values. Interestingly, the maximum luminosity detected at 6~GHz in two XRBs  with highly magnetized NSs is $\sim 10^{28}$~erg s$^{-1}$ regardless of the NS spin and magnetic field~\citep{2021MNRAS.507.3899V}. 

\section{Application to \sjs}\label{sec:application}
\phantom{}\sjs was the first Galactic source that exceeded by a factor of 10 the Eddington limit for a NS \citep{2018ApJ...863....9W}. It was therefore classified as a transient ULXP \citep[for a review, see][]{2017ARA&A..55..303K}. Its high luminosity and proximity made the source an ideal target for multi-wavelength monitoring, resulting in the first detection of radio emission from a BeXRB during an outburst~\citep{2018Natur.562..233V}. 

Gaia parallax measurements of \sjs indicate that the source is located at a minimum distance of 5~kpc~\citep{2018Natur.562..233V}. We therefore conservatively adopt this lower limit on the distance in our analysis.  \sjs has an almost circular orbit with orbital period of $\simeq 27.6$~d and projected semi-major axis $a_* \simeq 3.5\times10^{12}$~cm \citep{2018ApJ...863....9W}.

The system was observed at radio frequencies (at 4.5, 6, 7.5 and 22 GHz) at multiple epochs that covered a wide range of X-ray luminosities. Here, we adopt the 6~GHz and 22~GHz flux measurements of the system from \cite{2018Natur.562..233V, 2018MNRAS.473L.141V}. Whenever observations were performed at 4.5 and 7.5~GHz and a flat spectrum (i.e. $\alpha \approx 0$) was measured, then these fluxes were reported as representative of the 6~GHz flux~\citep{2018Natur.562..233V, 2018MNRAS.473L.141V}. The aforementioned papers provide also the $0.5-10$~keV energy flux at multiple epochs. To convert the latter flux to a bolometric X-ray  flux we adopt a bolometric correction factor $K_{\rm bol} \equiv F_{\rm bol}/F_{0.5-10 \, \rm keV} \approx 2.6$ from \cite{2018MNRAS.479L.134T}. Henceforth, we define periods with sub-Eddington and super-Eddington X-ray luminosities as epochs A and B, respectively. For completeness, the observed radio and X-ray luminosities for both epochs are listed in Table~\ref{tab:data}.

\begin{table}
	\centering
	\caption{Radio and X-ray luminosities of \sjs.}
	\label{tab:data}
	\begin{adjustbox}{width=1\columnwidth}
	\begin{threeparttable}
	\begin{tabular}{cccc} 
		\hline
Date (MJD) & $L_X$ ($10^{37}$ erg s$^{-1}$) &   $L_{6\, \rm GHz}$ ($10^{28}$ erg s$^{-1}$) &   $L_{22\, \rm GHz}$ ($10^{28}$ erg s$^{-1}$)\\

		\hline
		\multicolumn{4}{c}{Epoch A (Sub-Eddington)} \\
 58170  &  $1.03 \pm 0.02$  & $1.79 \pm 0.06$ &  $-$ \\
 58186  &  $0.47 \pm 0.02$ &  $0.44 \pm 0.07$ & $<0.76$ \\          
 58235  &  $0.058 \pm 0.003$ & $<0.25$ &  $-$\\ 
 58239  &  $0.43 \pm 0.09$ &   $<0.27$ & $-$ \\ 
 58242  &  $0.78 \pm 0.15$ & $1.17 \pm 0.11$ &  $-$\\ 
 58249  &  $1.5 \pm 0.2$  & $1.20\pm 0.12$ & $-$ \\
 58259  &  $0.39 \pm 0.08$ & $<0.24$ & $-$ \\
 \hline 
		\multicolumn{4}{c}{Epoch B (Super-Eddington)}   \\ 
58036		& $11.12 \pm 0.08$ & $<0.2$ & $<0.6$\\
58065	    & $287.0 \pm 1.9$ & $1.38 \pm 0.08$ &  $-$ \\
58072		& $184.3 \pm 1.2$ &  $1.66 \pm 0.07$ & $2.6 \pm 0.3$  \\
58078		& $124.4 \pm 0.9$ & $1.14 \pm 0.08$ & $1.9 \pm 0.4$  \\
58079		& $110.4 \pm 0.8$ & $0.99 \pm 0.08 $  & $2.0 \pm 0.5$  \\
58085		& $81.7 \pm 0.5$ &  $0.62 \pm 0.07$ & $2.0 \pm 0.3$  \\
58089		& $50.3 \pm 0.3$ &  $0.44 \pm 0.08$ & $1.8 \pm 0.3$ \\
58127		& $12.75 \pm 0.08$ & $0.38 \pm 0.07$ &  $-$\\
		\hline
	\end{tabular}
	\begin{tablenotes}
\item	Errors and upper limits are $1\sigma$ and $3\sigma$ respectively. Data are adopted from  \cite{2018Natur.562..233V, 2018MNRAS.473L.141V}.
	\end{tablenotes}
	\end{threeparttable}
	\end{adjustbox}
\end{table}

We then compare our toy model to the joint radio data (6 and 22 GHz) of each epoch separately. 
For the application to the data we will employ a Bayesian approach to ensure accurately estimated model parameters and their associated uncertainties.
Our goal is to infer the posterior
probability density $p$ given a dataset ($\mathcal{D}$) and priors from the Bayes' theorem for a model with a set of parameters contained in the vector $\theta$.
Having calculated the model we construct a likelihood function that can be defined as:
\begin{equation}
\ln{\mathcal{L}_i(\theta|\mathcal{D}_i)} =  -\frac{1}{2} \sum_{k}^{}\sum_{j}^{} \frac{(f_{\rm model,j}^k-f_{\rm data, j}^k)^2}{(\delta f_{\rm data, j}^k)^2}
\end{equation}
where $f_{\rm model, j}$ is the model radio flux at date $j$, $f_{\rm data, j}$ is the corresponding observed flux, $\delta f_{\rm data,j}$ is the associated error,  and the index $k$ runs over the different observing frequencies (here, 6 and 22 GHz).

We then derive posterior probability distributions for the model parameters and the Bayesian evidence with the nested sampling Monte Carlo algorithm
MLFriends \citep{2014arXiv1407.5459B, 2019PASP..131j8005B} using the
UltraNest\footnote{\url{https://johannesbuchner.github.io/UltraNest/}} package \citep{2021JOSS....6.3001B}.

\begin{figure}
    \includegraphics[width=0.47\textwidth]{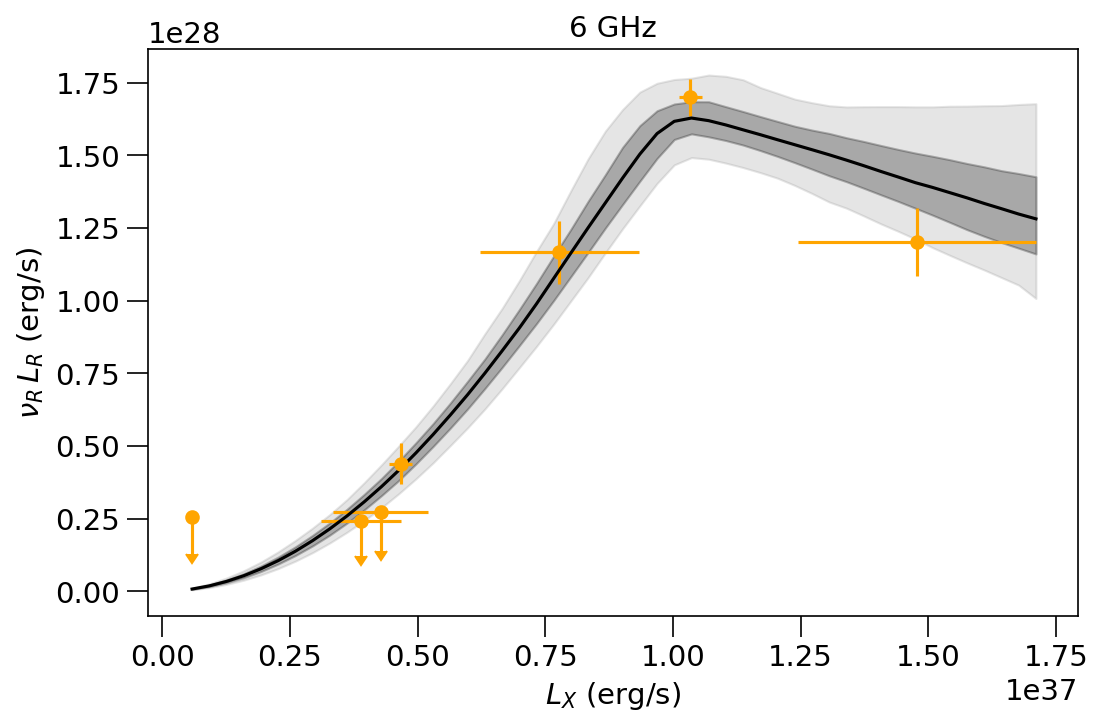}
    \includegraphics[width=0.47\textwidth]{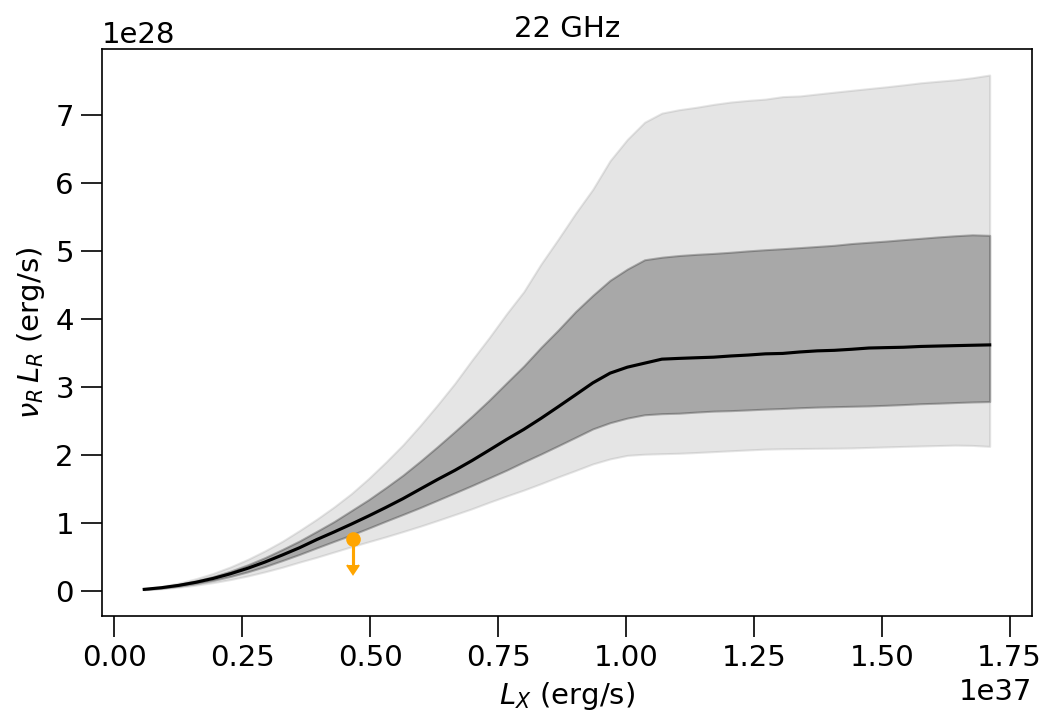}
    \includegraphics[width=0.47\textwidth]{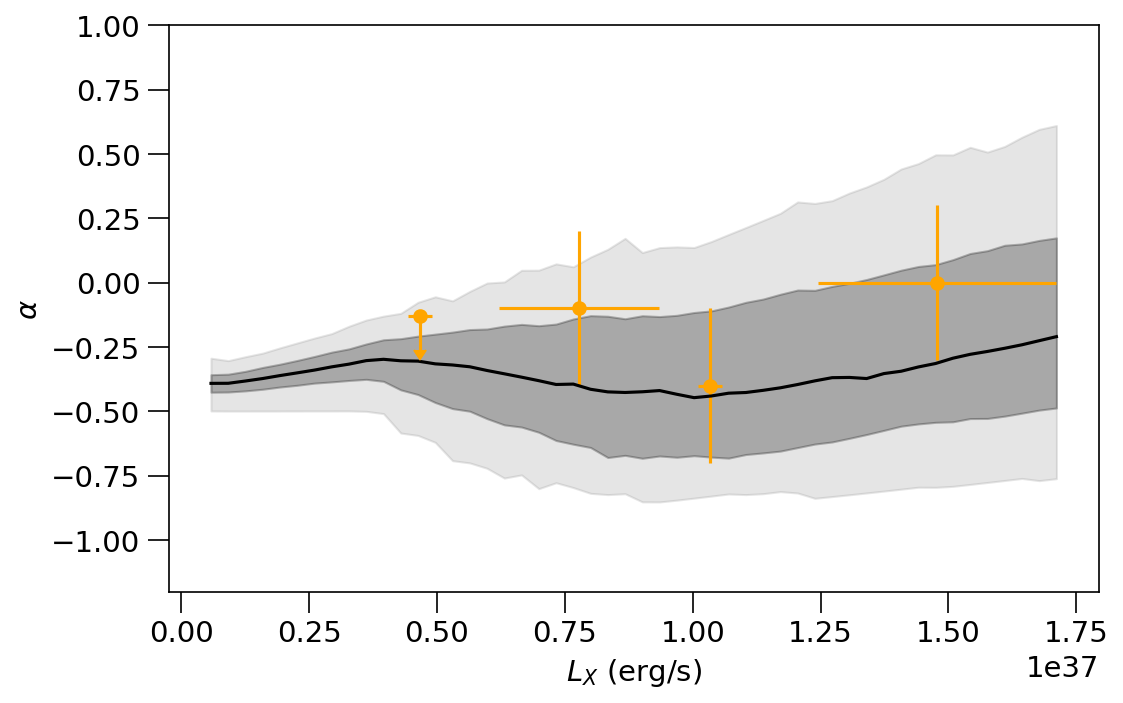}
    \caption{\textit{Top and middle panels:} Model fits to the 6~GHz and 22 GHz radio luminosities during epoch A. Dark and light shaded regions indicate the 68 per cent and 99 per cent confidence intervals, respectively.  \textit{Bottom panel:} Spectral index $\alpha$  between 6 and 22 GHz obtained from our model fit. Measured spectral indices are shown with symbols (adopted from \protect\cite{2018Natur.562..233V, 2018MNRAS.473L.141V}). Dark and light shaded regions indicate the 68 per cent and 99 per cent confidence intervals, respectively.}    \label{fig:fit-epA}
\end{figure}

\begin{figure}
    \centering
    \includegraphics[width=0.47\textwidth]{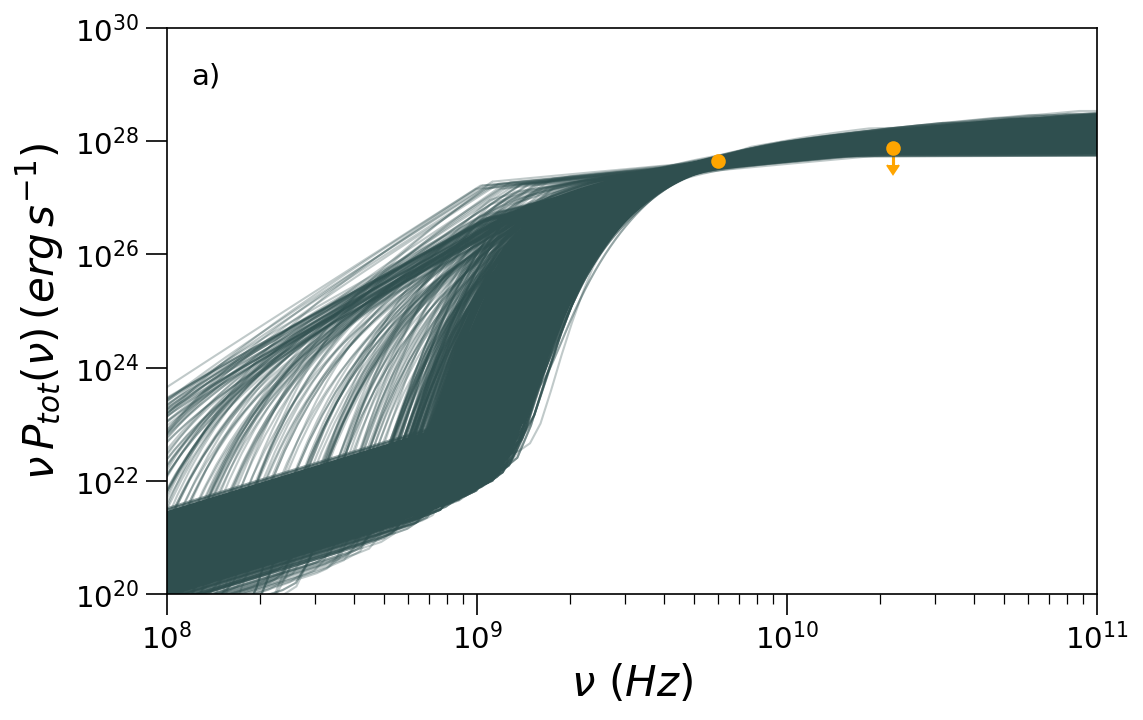}
    \includegraphics[width=0.47
    \textwidth]{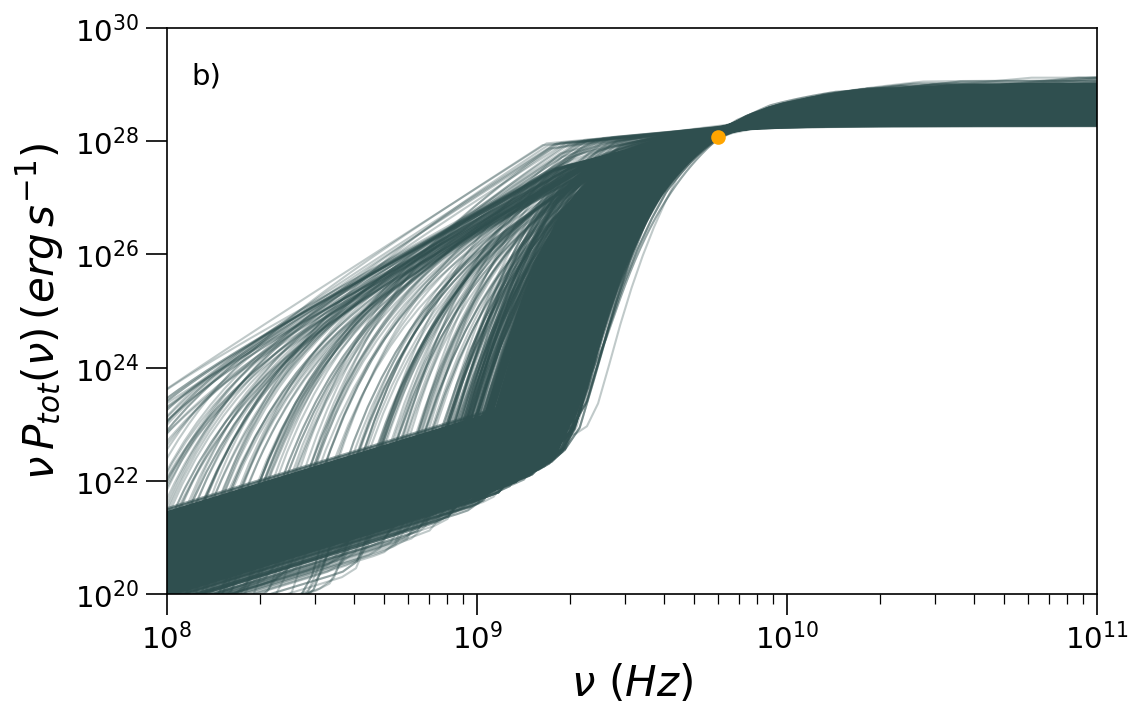}
    \caption{Radio spectra of \sjs computed using parameter values from the posterior distributions for two different dates from epoch A: 58186~MJD (panel a) and 58249~MJD (panel b).}
    \label{fig:spec-Sw}
\end{figure}

\begin{figure}
    \includegraphics[width=0.47\textwidth]{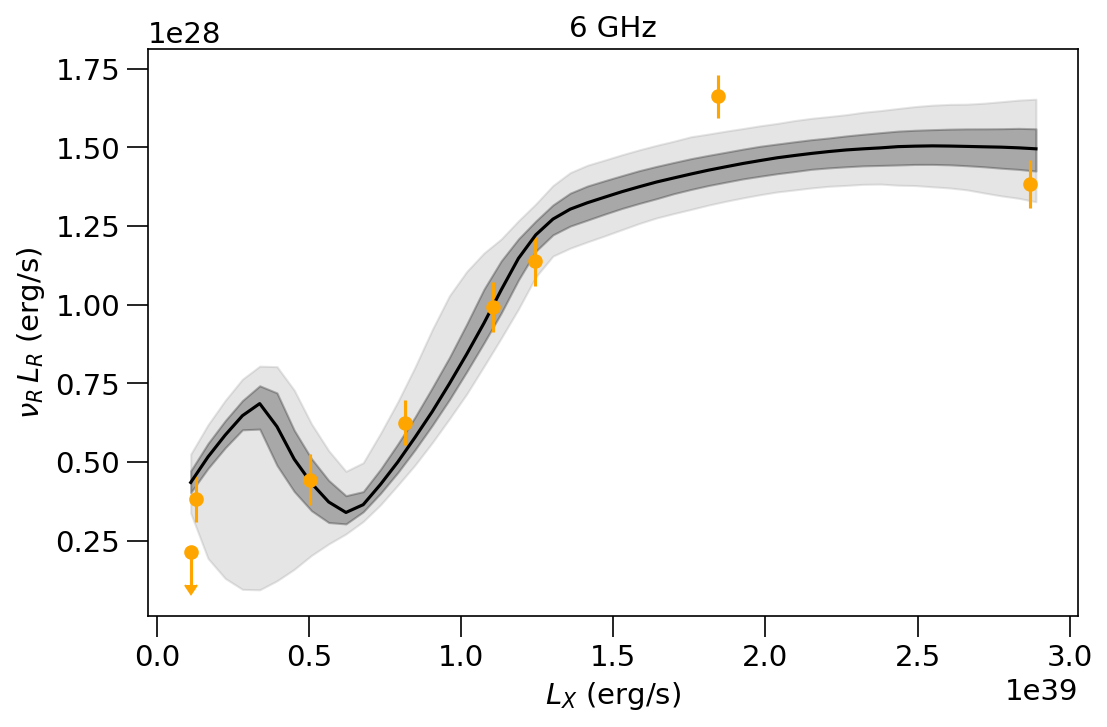}
    \includegraphics[width=0.47\textwidth]{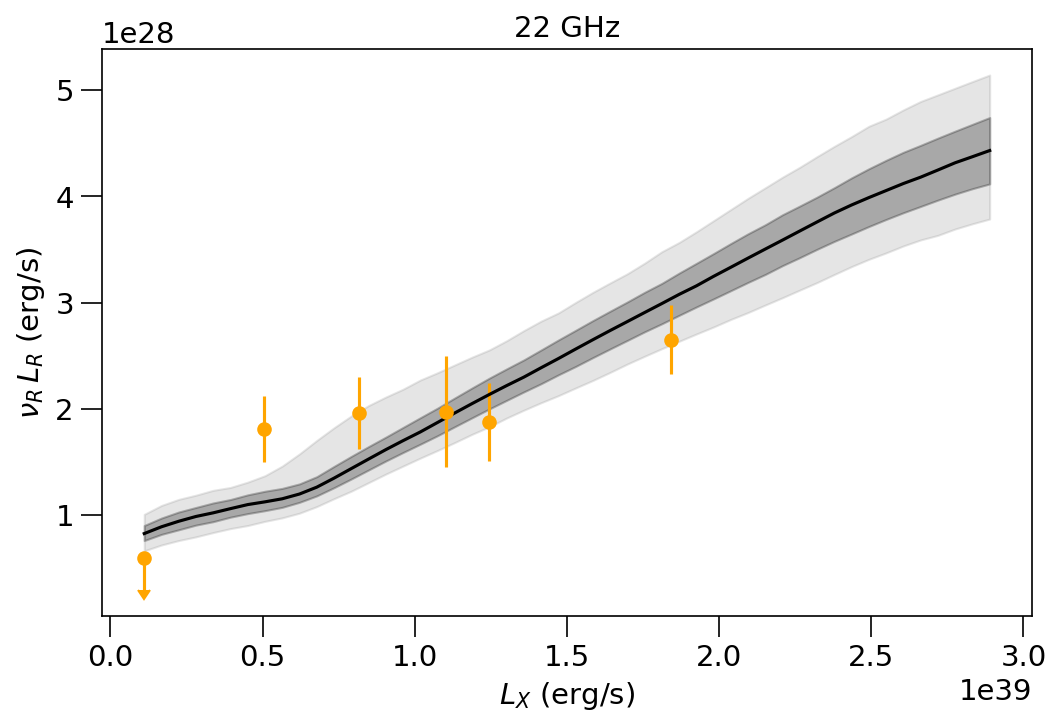}
    \includegraphics[width=0.47\textwidth]{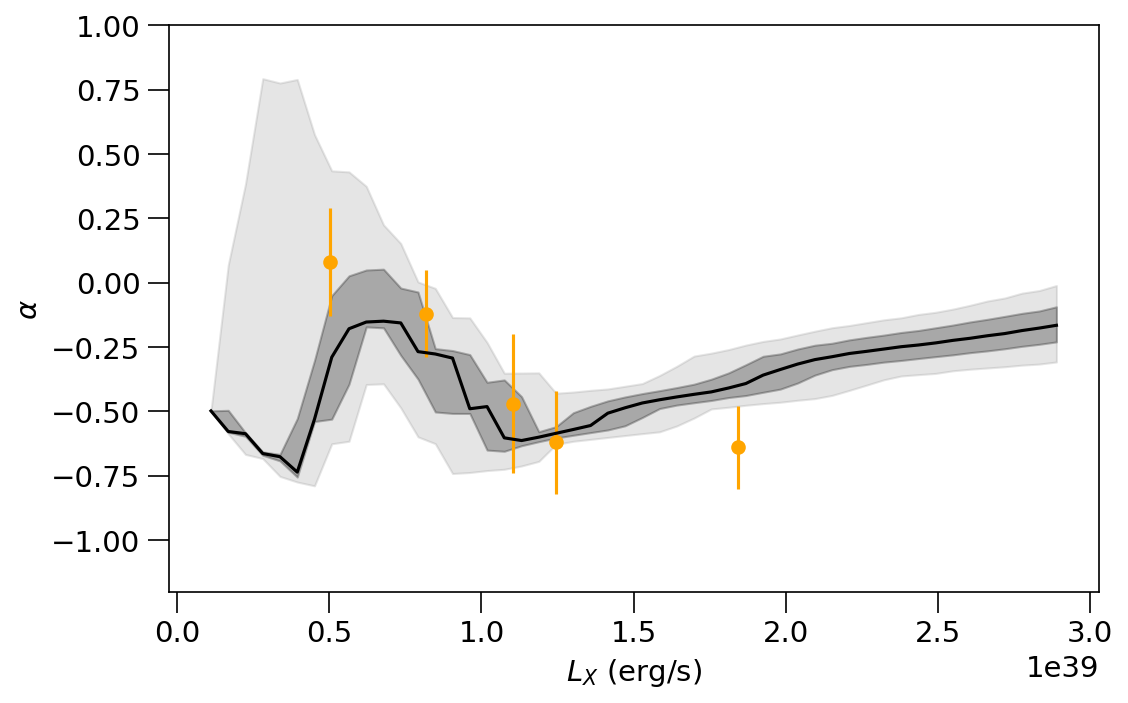}
    \caption{Same as in Fig.~\ref{fig:fit-epA} but for epoch B. All of the displayed solutions lead however to non-physical solutions with $d_{\rm Be} < 20 R_\odot$.}
    \label{fig:fit-epΒ}
\end{figure}

During this procedure we fix $B_{\rm NS}=10^{12}$~G, $R_{\rm NS}=10$~km, $M_{\rm NS}=1.4 M_{\odot}$, $\epsilon=0.2$, $\chi=2$ and $p=2$. We adopt $L_{\rm Be}=3\times10^4 \, L_{\odot}$ based on the spectral type of the companion star in \sjs \citep[O9.5V,][]{2020A&A...640A..35R}. For all other parameters (a complete list can be found in Table~\ref{tab:f_par}) we assume uniform prior distributions over a wide range of values except for the binary separation distance, $D$. For the latter parameter we consider a narrower range, i.e. $a_* \le D \le 2 a_*$, as even larger $D$ would imply a very massive companion ($\gtrsim 27 \, M_{\odot}$) which can be excluded based on spectroscopic observations of the system~\cite{2020A&A...640A..35R}.  We finally impose the constraints $\epsilon_{\rm e}=\epsilon_{\rm B}$ equipartition between relativistic particles and magnetic fields) and $i = \sin^{-1}(a_*/D)$.

The $L_R-L_X$ diagram for epoch A and the two radio frequencies is presented in Fig.~\ref{fig:fit-epA} (top and middle panels).  The radio data in epoch A can be satisfactorily explained by the model. The large uncertainty in the 22~GHz flux for $L_X \gtrsim 10^{37}$~erg s$^{-1}$ is also reflected to the spread in the values of the predicted spectral index in the $6-22$~GHz range (see bottom panel in Fig.~\ref{fig:fit-epA}). The observational values are computed assuming a power-law spectrum between two frequencies of reference. We adopt a similar approach when deriving the indices plotted in the figure, since the spectrum can be well described by a power law in the narrow range of 6 and 22 GHz most of the times~(see e.g. inset plots in Fig.~\ref{fig:RadioSpec37}). Because of the large uncertainties in the observational values of $\alpha$, a shock origin of the radio data (with $\alpha \le 0$) cannot be excluded at the $3\sigma$ level as shown in Fig.~\ref{fig:fit-epA}, in contrast to what has been argued by \cite{2018MNRAS.473L.141V}. 

The corresponding corner plot with the posterior probability distributions of the model parameters is shown in Fig.~\ref{fig:corner-epA}. The radio data of epoch A can be described well by the model 
for reasonable wind parameters ($\dot{M}_{\rm Be}\simeq3\times 10^{-9}~M_{\odot}$~yr$^{-1}$ and $V_{\rm Be}^\infty \simeq 1260$~km s$^{-1}$), if about 16 per cent of the accreted material to the NS is lost through outflows (see Fig.~\ref{fig:corner-epA}). Moreover, the minimum electron Lorentz factor and the fraction of shocked fluid energy transferred to particles and magnetic fields are well constrained (i.e. $\gamma_{\min}\approx 16$ and $\epsilon_{\rm B}\simeq 0.5$). The true binary separation distance is, however, not well determined as illustrated by the posterior distribution that is approximately uniform (as the adopted prior distribution). We also performed the simulations with $p$ being sampled from a uniform distribution between 2 and 4, but the power-law slope could not be constrained either. Hence, a more complex physical model where the particle injection spectrum would have had an angular dependence (e.g. softer further from the shock apex) would not be easily distinguished by a simpler one where $p$ is constant using the existing data.

For completeness we also show in Fig.~\ref{fig:spec-Sw} the model predicted  spectra for parameter values drawn from the posterior distributions for two different dates. Observational data are overplotted with symbols, whenever available (see also Table~\ref{tab:data}). Broadly speaking, we find that both observing radio frequencies fall into the optically thin part of the synchrotron spectrum, while the free-free emission of the stellar wind is always sub-dominant in agreement with the estimates of \cite{2018Natur.562..233V}.

Application of the model to epoch B shows that the posterior probability distributions do not differ much (within $1\sigma$) from those of epoch A, except for $\dot{M}_{\rm Be}$ and $V_{\rm Be}^\infty$ whose distributions  shift to smaller and higher values respectively (see Fig.~\ref{fig:corner-epB}). 
Moreover, the posterior distribution of the binary separation distance is not uniform as in epoch A, but shifted toward the upper end of the assumed range of values. In general, all posterior distributions are narrower than those in epoch A due to the larger number of observational constraints in epoch B.  Our toy model can satisfactorily describe the data, as shown in the $L_R-L_X$ diagrams at 6 and 22 GHz and the spectral index plot (see Fig.~\ref{fig:fit-epΒ}). However, none of the displayed models is physically meaningful, as they predict the formation of the WCR at distances less than $20 R_\odot$ from the Be star. In this regime, our assumption of a stellar wind moving with its terminal radial velocity breaks down. If the wind is moving with much lower speeds, then the formation of a strong shock might also be questionable. 

\section{Discussion}\label{sec:discuss}
We have presented a one-zone model for the radio emission from HMXB systems and computed the expected $L_R - L_X$ relation under several simplifying assumptions. In this section, we  discuss some of these aspects and their possible impact on our results. We also discuss the findings of our toy model in light of recent radio observations of a sample of NS X-ray binaries~\citep{2021MNRAS.507.3899V}.

\subsection{High-energy emission}
In our study we have not discussed any potential signals of high-energy non-thermal radiation from HMXBs. The massive companion star is a source of low-energy radiation that can be up-scattered to high energies by the radio-emitting electrons in the WCR. In fact, for typical stellar luminosities and magnetic field strengths in the WCR, electrons lose energy through inverse Compton scattering more efficiently than synchrotron radiation (see also section~\ref{sec:radiation}).  Electrons radiating at 10~GHz have Lorentz factors $\gamma_R \sim 100$, assuming typical parameter values (i.e., $\dot{M}_{\rm Be}=10^{-9}\, M_{\odot}$~yr$^{-1}$, $V_{\rm Be}^\infty=10^3$~km s$^{-1}$, $\epsilon_{\rm B}=0.1$, $\eta_{\rm out}=0.1$). These electrons can inverse Compton scatter stellar optical photons at X-ray energies.  The ratio of inverse Compton  and synchrotron luminosities produced by radio-emitting electrons can be approximately written as $(\nu L_{\nu}|_{\rm IC})/(\nu L_{\nu}|_{\rm syn}) \approx U_{\rm ph}/U_{\rm B} = 8 L_{\rm Be}/(9 c \, \epsilon_{\rm B} \dot{M}_{\rm Be} V_{\rm Be}^\infty) \simeq 1800\, L_{\rm Be, 4} \epsilon_{\rm B, -1}^{-1} \dot{M}_{\rm Be, -9}^{-1} (V_{\rm Be, 3}^{\infty})^{-1}$, where $L_{\rm Be, 4}=10^4 L_{\odot}$ and Thomson scattering was assumed. The only two HMXBs hosting strongly magnetized NSs have GHz radio luminosities $\sim 10^{28}$ erg~s$^{-1}$~\citep{2021MNRAS.507.3899V}. Assuming that their radio emission is powered by synchrotron radiation from the WCR, the predicted X-ray luminosity due to inverse Compton scattering cannot be much higher than $\sim 10^{32}$~erg s$^{-1}$. This prediction is still many orders of magnitude below the X-ray luminosity produced via accretion onto the NS, hence it cannot be used as a probe of the physical conditions in the radio-production site of the system. The inverse Compton scattered emission can extend and actually peak (in $\nu L_\nu$ units) in the gamma-ray band, as shown for colliding wind binary systems~\cite[e.g.][]{2006ApJ...644.1118R, 2013A&A...555A.102W, 2021MNRAS.504.4204P}. Even though many of these systems have been observed in radio wavelengths~\citep[for a catalog, see][]{2013A&A...558A..28D}, the predicted non-thermal high-energy emission in X-rays and gamma-rays remains elusive. So far, non-thermal X-rays and TeV gamma-rays have been detected only from $\eta$ Carinae~\citep{2018NatAs...2..731H, 2020A&A...635A.167H}. Detailed radiative models for high-energy emission from colliding wind binaries show that the GeV gamma-ray luminosity from inverse Compton scattering can be up to $10^2-10^4$ times higher than the GHz radio luminosity, depending on the binary separation distance and the orbital phase~\citep[e.g.][]{2021MNRAS.504.4204P}. Similar ratios can be expected in our scenario, as it bears many similarities with the shock-wind models for massive binaries. However, due to the lower radio luminosities of HMXBs with magnetized NSs~\citep{2021MNRAS.507.3899V} any gamma-ray signal would be even more difficult to detect than in massive binaries. 

\subsection{Simplifying assumptions}
In our analysis we considered $\eta_{\rm out}$ constant, but in principle it can depend on the accretion rate onto the NS. For instance, if the NS outflows are driven by changes in the disc structure as the accretion rate increases, then one would expect $\eta_{\rm out} \ll 1$ for $L_X \ll L_{\rm Edd}$ and $\eta_{\rm out} \sim 0.1-0.5$ otherwise~\citep[e.g.][]{2019MNRAS.484..687M}. How would such an $L_X$-dependent outflow fraction change the $L_R-L_X$ diagram? This can be qualitatively understood by looking at panel d in Fig.~\ref{fig:multiple}. A decrease in $\eta_{\rm out}$ would not affect the maximum predicted radio luminosity, but would require higher $L_X$ values in order to achieve it. If the outflow fraction changed from 0.005 to 0.5 at  $L_{X} \sim 10^{38}$ erg~s$^{-1}$, then the $L_R-L_X$ diagram would be the combination of the blue and brown coloured curves. In this scenario, it is therefore not possible to explain similar radio luminosities for sub-Eddington and super-Eddington X-ray luminosities, as observed in \sjs~\citep{2018MNRAS.474L..91V}. 
 On the contrary, we found that an almost constant outflow fraction $\eta_{\rm out}$ and a variable Be stellar wind could broadly describe the multi-epoch observations of \sjs (see e.g. Figs.~\ref{fig:corner-epA} and \ref{fig:corner-epB}.

Effects on the structure of the WCR due to the orbital motion were not  taken into account. If the orbital velocity is sufficiently large (i.e. close binaries), then the WCR would wrap around the stars because of the Coriolis forces acting upon it. The structure of the WCR in the orbital plane would look like a spiral~\citep[e.g.][]{2007ApJ...662..582L, 2008MNRAS.388.1047P, 2015A&A...577A.122J, 2020ApJ...900L...3R}. Hence, the line of sight would cut through more regions containing non-thermal particles. If the radio-emitting electrons that are injected into the WCR close to the apex do not suffer strong energy losses as they stream to the more distant regions of the WCR, then they could contribute to the  synchrotron radio luminosity of the system. An orbital modulation of the radio flux would be expected in this case. The ratio of the orbital velocity to the wind speed is a useful estimator of the importance of orbital motion effects~\citep{2007ApJ...662..582L}. Assuming an almost circular orbit, like in \sjs, the ratio reads $V_{\rm orb} / V^{\infty}_{\rm Be} = \sqrt{G (M_{\rm NS}+M_{\rm Be})/D} / V_{\rm Be}^\infty \simeq 0.06\, (D/10^3 R_\odot)^{1/2} (V_{\rm Be}^\infty/10^3 \, {\rm km \, s}^{-1})^{-1}$, where a $20 M_{\odot}$ companion star was assumed. Thus, for typical parameters used in this study, orbital effects should not affect the shape of the WCR.
 
In this proof-of-concept work we adopted a one-zone toy model for the calculation of the synchrotron radiation from the WCR. We assumed that all physical quantities (e.g., post-flow velocity, post-shock energy density, electron power-law slope and others) remain constant in the emitting volume, having values equal to those computed at the shock apex. Because the shape and properties of WCR change at large angles from the line of centers \citep[e.g.][]{2020MNRAS.495.2205P}, we limited our calculations to a spherical wedge with opening angle $\pi/4$ (with respect to the line of centers), corresponding to a volume $\pi R_{\rm sh}^3$.  A more accurate calculation would require a multi-zone radiation model that follows the particle distribution along the streamlines of the post-shock fluids, taking into account radiative and adiabatic energy losses~\cite[e.g.][]{2016A&A...591A.139D,2017A&A...598A..13D, 2020ApJ...904...91V, 2021MNRAS.504.4204P}. Such multi-zone models can also account for potential changes in the particle injection spectrum (e.g. power-law slope and accelerated energy fraction) as a function of angle due to changes in the shock obliquity and Mach number~\citep[see][and references therein]{2021MNRAS.504.4204P}. Angular-dependence in the physical parameters of the model would generally lead to smoother synchrotron spectra than those presented here, unless the total spectrum is dominated by particles within a small range of angles \citep[see e.g. Fig.~3][]{2016A&A...591A.139D}. 
 Another effect that cannot be captured by a one-zone model for radio emission is that the free-free opacity may vary along the line of sight, if the synchrotron emitting region is physically extended, as observed in certain radio-emitting colliding-wind binaries~\citep[see][and references therein]{2003A&A...409..217D}. Nevertheless, the one-zone emitter may still be a good approximation for NS XRBs, like \sjs, that are detected as point-sources in radio frequencies~\citep{2018Natur.562..233V, 2018MNRAS.474L..91V}. In fact, \sjs is about 10 times more distant with $\sim 10^{-3}$ times lower flux in GHz frequencies than WR~147, a well-studied colliding wind massive binary systems with resolved radio emission~\citep[e.g.][]{2002MNRAS.334..631W, 2003A&A...409..217D, 2020A&A...640A..35R}. 

\subsection{Interpretation of modelling results of \sj}
Using a Bayesian approach we searched for model parameters that describe the data of \sjs at 6 and 22 GHz over a wide range of X-ray luminosities, and found that our toy model can describe reasonably well the data. The posterior distributions of the model parameters found for epoch A are physically meaningful. No significant changes in the model parameters were required for explaining the data of epoch B except for the properties of the stellar wind: the mass loss rate  decreased by a factor of $\sim 25$, while its terminal velocity increased by a factor of $\sim 4$ (see Figs.~\ref{fig:corner-epA} and \ref{fig:corner-epB}). These values result in a lower wind momentum, which when combined with the higher X-ray luminosities, translate to a very weak stellar wind. If accretion onto the NS took place through the wind, then the combination of the above mentioned effects should reduce the mass accretion rate onto the NS. It would be then difficult to explain the high accretion rates sustained for $\sim3$-month period in epoch B. However, in BeXRBs mass transfer is due to interactions of the NS with the Be disc, a mechanism that seems to hold even in major outbursts that last for several orbital periods~\citep[e.g.][]{2014ApJ...790L..34M}. Thus, the NS can accrete material that has been already captured by the NS at a prior time, and a WCR form because of the NS outflow  colliding with the stellar wind. To better understand the dynamics of this complex system of accretion inflow/outflow and stellar wind/disc three-dimensional relativistic (magneto-)hydrodynamical simulations should be performed.

The derived stellar wind properties for epoch B are also in tension with theoretical and numerical studies of hot stellar winds~\citep[e.g.][]{1980A&A....87..102F, 1994ApJ...435..756B, 2015A&A...579A.111K}. For high enough X-ray luminosities the gas that resides on the stellar surface exposed to the X-ray source becomes highly photoionized, thus suppressing the wind velocity or even halting the formation of a radiatively driven wind from that side. The effect of the X-rays on the stellar wind is expected to be stronger for larger X-ray luminosities, lower mass loss rates, and faster winds~\citep[e.g.][]{2015A&A...579A.111K}, as found for epoch B. More recently \cite{2018A&A...620A.150K} showed that clumping of the Be stellar wind in the radial direction can weaken the effects of X-ray irradiation because of the balancing effect of recombination. Even if clumpiness is not taken into account, stellar wind can still escape from the other side of the star, causing the so-called `shadow wind' that is often seen in luminous HMXBs \citep[as in e.g. 4U\,1700-37,][]{1989ApJ...343..409H}. In fact, the shadow wind model \citep{1994ApJ...435..756B} has been offered as a possible mechanism to explain periodic dips (on orbital timescales) seen in the ULXP M51 ULX-7 \citep{2021ApJ...909...50V} and in SMC X-2 during its 2015 super-Eddington outburst~\citep{2016ApJ...828...74L}. Summarizing, X-ray emission from the compact object can lead to anisotropic and inhomogeneous stellar winds. However, these effects on the WCR emission are beyond the scope of this work.

\subsection{Jets in HMXBs?} 

We demonstrated that a shock origin of the radio emission of \sjs for $L_X < L_{\rm Edd}$ is possible with reasonable parameter values, while it is disfavoured for the super-Eddington accretion phase of the outburst based on physical arguments. Still, the flat radio spectrum ($\alpha\le 0$) observed in this phase (see Fig.~\ref{fig:fit-epΒ}) implies the presence of non-thermal emitting particles. This raises the question of where these particle could be found besides the WCR. While radio-emitting jets (i.e. collimated outflows) are ubiquitous in low-mass X-ray binaries with weakly magnetized NSs ($B_{\rm NS}\le 10^9$~G), they have not been unambiguously identified in HMXBs with strongly magnetized NSs. The lack of solid observational evidence for the presence of jets in combination with the incomplete theoretical understanding of jet launching in such systems makes the interpretation of data within the jet scenario dubious. For instance, \cite{2012A&A...538A...5K} argued that no compact jet can be formed in accreting NSs with $B_{\rm NS} \gtrsim 10^{12}$~G, because the physical mechanism producing large-scale magnetic fields in the disc needed for the jet formation cannot operate efficiently. \cite{2008A&A...477....1M} also argued that no jet can form in accreting NSs with $B_{\rm NS}>10^{12}$~G because the strong magnetic field cannot be twisted, as it is always dynamically important (compared to the gas in the inner disc). Alternatively, a jet could be formed by collimating the pulsar wind moving along opened field lines. Collimation, which could be mediated by external large-scale magnetic fields or by the pressure of dense plasma (e.g. from disc winds), would be necessary to make the pulsar wind appear as a typical radio jet, as in their large-scale analogs in active galaxies \citep[e.g.][]{2011MNRAS.418L..79T, 2015ASSL..414..177V, 2021A&A...647A..67B}.  Recently, \cite{2016ApJ...822...33P} studied the disc-induced opening of magnetic field lines in accreting millisecond pulsars, and derived a measure of the power carried by the pulsar wind as a function of the mass accretion rate onto the NS, namely $L_j \propto \dot{M}^{4/7}$. While this does not relate directly the radio luminosity, which depends on the uncertain nature and location of the dissipation mechanism in the jet as well as the jet bulk motion, with $L_X$ it is useful for our qualitative discussion. Given that radiation-driven outflows from the disc are expected for super-Eddington accretion rates~\citep{1973A&A....24..337S, 2007MNRAS.377.1187P, 2018A&A...610A..46C}, the required collimation of the pulsar wind might be possible in \sjs for $L_X > L_{\rm Edd}$. In epoch A ($L_X < L_{\rm Edd}$) the much lower accretion rates would power a weaker jet even if the collimation was still present. Hence, a hybrid scenario where the radio emission is dominated by the WCR at sub-Eddington X-ray luminosities and by a collimated outflow (jet) at super-Eddington luminosities is plausible.

The collimated pulsar wind model of \cite{2016ApJ...822...33P} was also discussed in the light of recent radio observations of HMXBs with strongly magnetized NSs~\citep{2021MNRAS.507.3899V}. These authors found no correlation between radio and X-ray luminosity for the sources in the sample (and for \sjs individually) in contrast to the scaling $L_j \propto \dot{M}^{4/7}$. Moreover, the radio luminosity of the sources in the sample did not seem to depend on the NS magnetic field and/or spin, as theoretically predicted. These findings imply that either the jet power and radio luminosity have a complicated relation that depends on the accretion rate, or that more than one radio production regions might be at work in HMXBs with strongly magnetized NSs. 

\subsection{Potential diagnostics for the radio origin}
The above discussions raise an interesting question: would it be possible to discriminate between a jet and a shock origin for the radio emission in HMXBs? Potential diagnostics include the (i) radio/X-ray correlation, (ii) radio/NIR/optical polarization, (iii) orbital modulation of radio fluxes, and (iv) radio-to-NIR spectral shape. For the sake of argument, let us assume that a compact steady jet can be formed in XRBs with highly magnetized NSs as in the low-hard states of BH XRBs. We can therefore compare our model predictions with those of well-tested jet models for the the low-hard states of BH XRBs \citep[e.g.][]{2003MNRAS.343L..59H, 2003MNRAS.345.1057M}. 
\begin{enumerate}[label=(\roman*)]
    \item Even though our model makes clear predictions about the radio/X-ray correlation (see e.g. equation~\ref{eq:LrLx} and Figs.~\ref{fig:LCex}-\ref{fig:multiple}), similar shapes can be obtained in the context of compact jet models. For instance, jet models accounting for both the radio and X-ray emissions in BH XRBs predict $L_R \propto L_X^{1.4}$ for a wide range of $L_X < L_{\rm Edd}$~\citep[e.g.][]{2003A&A...397..645M}. Depending on the stellar wind properties, observations of HMXBs could capture only the steep branch of the correlation in our scenario, which has a similar slope ($L_R \propto L_X^{1.7})$. Only through quasi-simultaneous radio/X-ray observations of individual systems mapping a wide range of mass accretion rates up to the Eddington limit, as in \sjs, would be possible to critically test the jet and shock models for radio emission. 
    \item Radio polarization might be not ideal for distinguishing between a jet and shock origin of the radio emission. In both cases, radio is produced by synchrotron radiation, whose  polarization degree (PD) depends on the magnetic field geometry and properties of the external medium. In both models, there is a degree of freedom in regard to the magnetic field configuration in the emission region. For instance, a low PD in a jet scenario would be expected if the plasma is turbulent in the radio emitting region, while a higher PD would be expected if the magnetic field is predominantly toroidal, as is theoretically predicted in the inner regions of magnetically dominated jets \citep[e.g.][]{2001Sci...291...84M, 2004ApJ...605..656V}. Similarly, in the WCR the magnetic field will depend on the binary separation distance, the magnetic field of the stellar wind, and the development of turbulence in the shock region~\citep[see][and references therein]{2021MNRAS.504.4204P}. Variable PDs are also expected in both scenarios, as they can be related to dynamical changes in the emission region. Synchrotron radiation on long wavelengths can also be depolarized by radiative transfer effects, such as Faraday rotation due to the thermal plasma of the stellar wind~\citep[][]{2017A&A...598A..42H}. As a result, a low or high PD cannot uniquely determine the origin of the radio emission. Meanwhile, polarization at NIR/optical frequencies cannot probe the non-thermal synchrotron emission in HMXBs, as it dominated by the stellar radiation of the massive companion (see Fig.~\ref{fig:composite_spec}). 
    \item Orbital modulation of the radio fluxes produced in the WCR or jet can be caused by variable attenuation from the wind of the massive star. Orbital modulation effects will depend on the inclination angle, binary separation distance and stellar wind properties (i.e.  temperature, mass loss rate, temperature, and clumpiness). \cite{2012MNRAS.422.1750Z} presented analytical expressions for the modulation depth due to free-free absorption by an isothermal wind when radio emission takes place close to the jet base ($z_0 \rightarrow 0$) or far away ($z_0 \gg D$). This analysis can also be applied to our scenario, if we neglect the curvature of the WCR (see Fig.~\ref{fig:ff-geo}). Because we consider radio emission from a small spherical wedge of the WCR, the corresponding $z_0$ cannot be much larger than the separation distance.  In a jet model, the radio spectrum is a superposition of emission from different distances along the jet \citep{1979ApJ...232...34B}, while in our model there is a single region producing the radio emission. Therefore, in a jet model a frequency-dependent modulation in flux is expected because of the different locations of radio production in addition to the frequency-dependence of the free-free absorption. A more careful analysis of these effects in the shock scenario is worth pursuing. From an observational point of view, HMXBs with day-long orbital periods would be ideal for the search of orbital modulation effects, since the X-ray luminosity can be considered constant within a period (in contrast to \sjs).
    \item Perhaps the most straightforward way of testing the jet and shock scenarios for radio emission is the shape of the spectrum from radio to NIR frequencies.  The radio spectrum of a compact jet usually shows a break close to NIR frequencies, above which it becomes optically thin~\citep[e.g.][]{2013MNRAS.429..815R}. Below that frequency the jet spectrum is $F_{\nu}\propto\nu^0$. The synchrotron self-absorbed part of the spectrum in a single-zone model like ours is $F_{\nu}\propto \nu^{2-2.5}$. Moreover, the transition to optically thin emission takes place usually at lower frequencies (0.1-10 GHz) than in the standard jet model -- see e.g. Figs~\ref{fig:SectraDecomp}-\ref{fig:RadioSpec39}. Quasi-simultaneous observations of individual systems performed over a wide range of frequencies ($\sim0.1$ GHz to $\sim100$~THz) would be important for identifying potential spectral breaks in the spectrum. 
\end{enumerate}

\subsection{Our results in a broader context}
\begin{figure}
    \centering
    \includegraphics[width=0.47\textwidth]{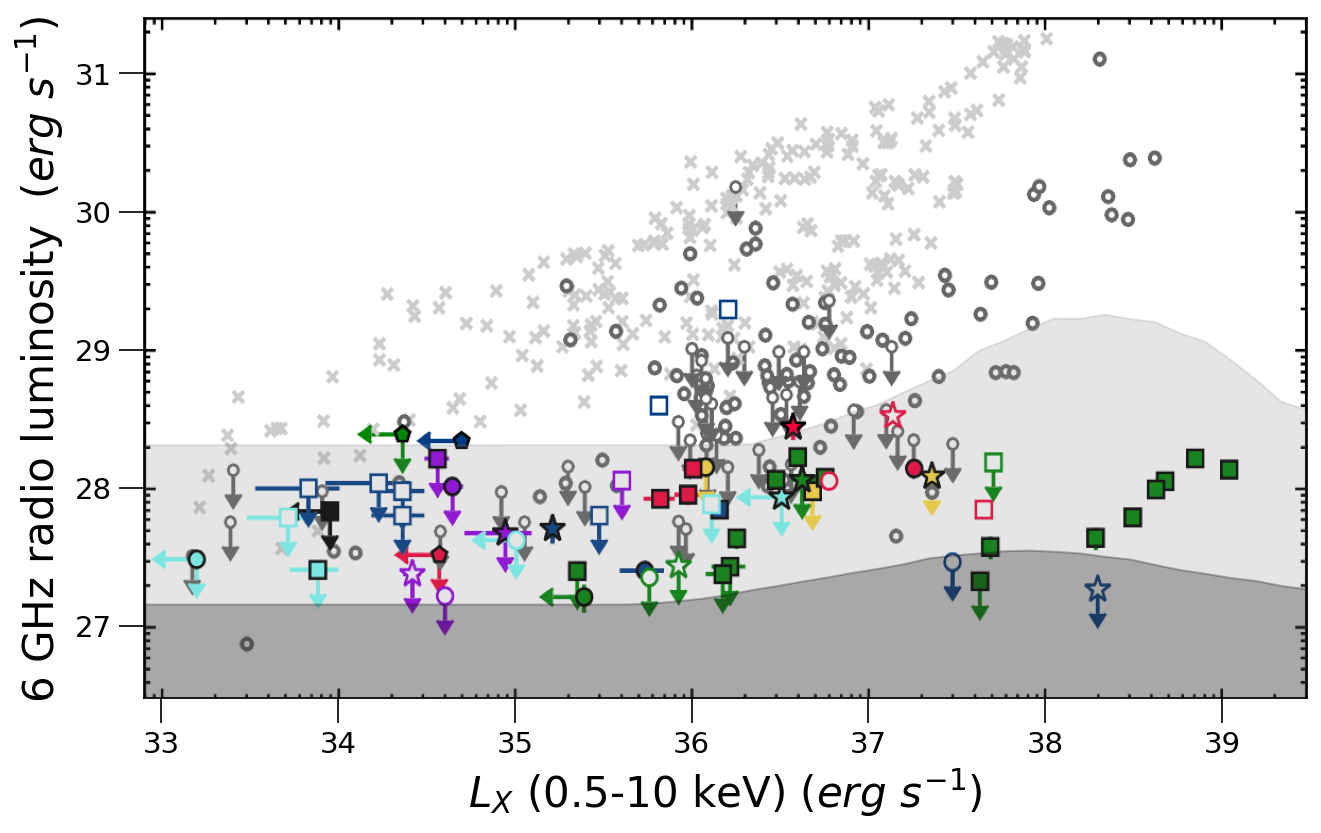}
    \caption{The $L_R-L_X$ plane for X-ray binaries. Figure adopted from \protect\cite{2021MNRAS.507.3899V}. Archival observations are shown in grey (crosses: black holes, circles: weakly magnetized NSs). Coloured symbols show recent observations of NS XRBs (filled markers: strongly-magnetized NSs, open markers: weakly magnetized NSs). The borders of the dark and light shaded regions mark respectively the 80th and 99th percentiles of our solutions (the full bands are not shown).}
    \label{fig:LrLx}
\end{figure}

As this is, to the best of our knowledge, the first quantitative study of radio emission from such systems, we summarize our predictions on the radio - X-ray luminosity plane for XRBs \citep[adopted from][]{2021MNRAS.507.3899V}. With coloured filled symbols are indicated HMXBs with strongly magnetized NS, including \sjs (green squares) and GX~1+4 (red squares). The dark and light shaded regions show the 68 per cent and 99 per cent range of our model predictions for 10,000 samples drawn randomly from the following uniform distributions\footnote{We have not excluded the non-physical solutions that might exist above a certain $L_X$ for different parameter values.}: $\log{\dot{M}}_{\rm Be} (M_\odot/{\rm yr}) \in [-9,-6]$, $\log(V_{\rm Be}^\infty) {\rm (km/s)} \in [2,3.5]$, $\log{\gamma_{\min}} \in [0.2,2.2]$, $\log(D) (R_{\odot}) \in [2,3]$, $\log(\eta_{\rm out}) \in [-2, -0.3]$, $\log(\epsilon_{\rm B}) \in [-3,0]$. All other parameters are fixed to their values listed in Table~\ref{tab:f_par}. We also assumed the same bolometric correction factor as in \sjs (see section~\ref{sec:application}) for estimating the 0.5-10 keV luminosity.  While the exact shape of the bands will depend on the adopted distributions, the flat part for $L_X \lesssim 10^{36}$~erg s$^{-1}$ implies that most solutions are dominated by the free-free emission of the wind in this range of X-ray luminosities.  When all physical parameters are randomly combined, then most of our solutions result in very low radio luminosities, i.e. $\sim 10^{22}-10^{24}$~erg s$^{-1}$ (well outside the plotting range of Fig.~\ref{fig:LrLx}), with the median being $10^{26}$~erg s$^{-1}$. To reach luminosities as high as $10^{28}$~erg s$^{-1}$ special parameter combinations are needed, as also suggested by our analysis in section~\ref{sec:results} (see also eq.~\ref{eq:Lxcr}). The borders of the dark and light shaded regions mark respectively the 80th and 99th percentiles of our solutions, where most of the upper limits and detections of NS XRBs are found. This means that, if our model is correct, we are just seeing the tip of the iceberg. 

\section{Summary and Conclusions}\label{sec:conclusions}
We have presented a toy model for radio emission in HMXBs with strongly magnetized NSs. We postulated that material is lost in the form of outflows from the accreting NS. The NS outflow, moving at a speed comparable to the Keplerian velocity at the inner disc radius, collides with the stellar wind of the massive companion forming a WCR. Radio emission is then expected from the system as a result of synchrotron radiation by shock-accelerated electrons in the WCR and free-free emission of the stellar wind. We explored the relation between the GHz luminosity and the X-ray luminosity powered by accretion for different model parameters and applied the model to \sjs, a unique BeXRB with multiple radio detections extending over a wide range in X-ray luminosities.

We generally find that the radio luminosity at GHz frequencies is written as $L_R \propto L_X^b$. No correlation with X-rays is expected ($b=0$) when the radio luminosity is dominated by the thermal emission of the stellar wind. Typically, for sub-Eddington X-ray luminosities we predict a steep correlation ($b\sim 12/7$) and a more shallow one ($b= 2(p-1)/7$) for super-Eddington X-ray luminosities, where $p$ is the power-law index of accelerated electrons. The exact transition happens at a critical X-ray luminosity, $L_{X, \rm br}$, which mostly depends on the momentum ratio of the two outflows. Finally, a very steep anti-correlation is expected when the free-free opacity at GHz radio frequencies is very high, which is more common for compact binaries. The maximum radio luminosity (when this is dominated by optically thin synchrotron emission) is independent of the NS properties, such as magnetic field, fraction of mass lost through outflows, and radiative efficiency of accreting material. It depends mostly on the stellar wind momentum, binary separation distance, and the minimum Lorentz factor of accelerated electrons.

When applied to \sjs, the model provides a good description of radio data for sub-Eddington X-ray luminosities with reasonable physical parameters. While the model can also fit the observations in the super-Eddington phase of the outburst, the parameters needed cannot be  physically justified. Moreover, the WCR is pushed so close to the companion star that several of our model assumptions break down. It is therefore plausible that the radio emission of \sjs is dominated by the WCR at sub-Eddington X-ray luminosities and by a collimated outflow (jet) at super-Eddington luminosities.

In conclusion, the multi-epoch observations of \sjs offer the first opportunity to quantitatively test theoretical scenarios for radio emission in HMXBs with strongly magnetized NSs and help us understand the coupling of disc winds and jets. Regular radio monitoring of \sjs and GX~1+4 during different levels of X-ray luminosity and searches for orbital modulation in radio fluxes might help us distinguish between different scenarios. Looking into the future, more sensitive instruments like ngVLA (with ten times the sensitivity of JVLA and ALMA) \citep{2018SPIE10700E..1OS} might reveal a large number of HMXBs with strongly magnetized NSs, constrain the slope of the $L_R-L_X$ relation, and put our model into test. 

\section*{Acknowledgements}
The authors would like to thank the anonymous referee for insightful comments on the manuscript, Dr. J. Buchner for his advice on the use of the UltraNest package and its implementation and Prof. N. Kylafis for comments on the manuscript. M.P. would like to thank the Observatory of Strasbourg for its hospitality during her visit in which this paper was completed. M.P. acknowledges support from the MERAC Foundation through the project THRILL. 
 
This research made use of Astropy,\footnote{http://www.astropy.org} a community-developed core Python package for Astronomy \citep{astropy:2013, astropy:2018}, and UltraNest software package,\footnote{\url{https://johannesbuchner.github.io/UltraNest/}} for model-to-data comparison using nested sampling \citep{2021JOSS....6.3001B}.
 

\section*{Data Availability}
Data used in this work are publicly available 
through published papers \citep{2018MNRAS.474L..91V, 2018Natur.562..233V,2021MNRAS.507.3899V}.

\bibliographystyle{mnras}
\bibliography{swiftJ0243} 

\appendix
\section{Optical depth for free-free absorption by the stellar wind}\label{app:tff}

\begin{figure}
    \includegraphics[width=0.49\textwidth]{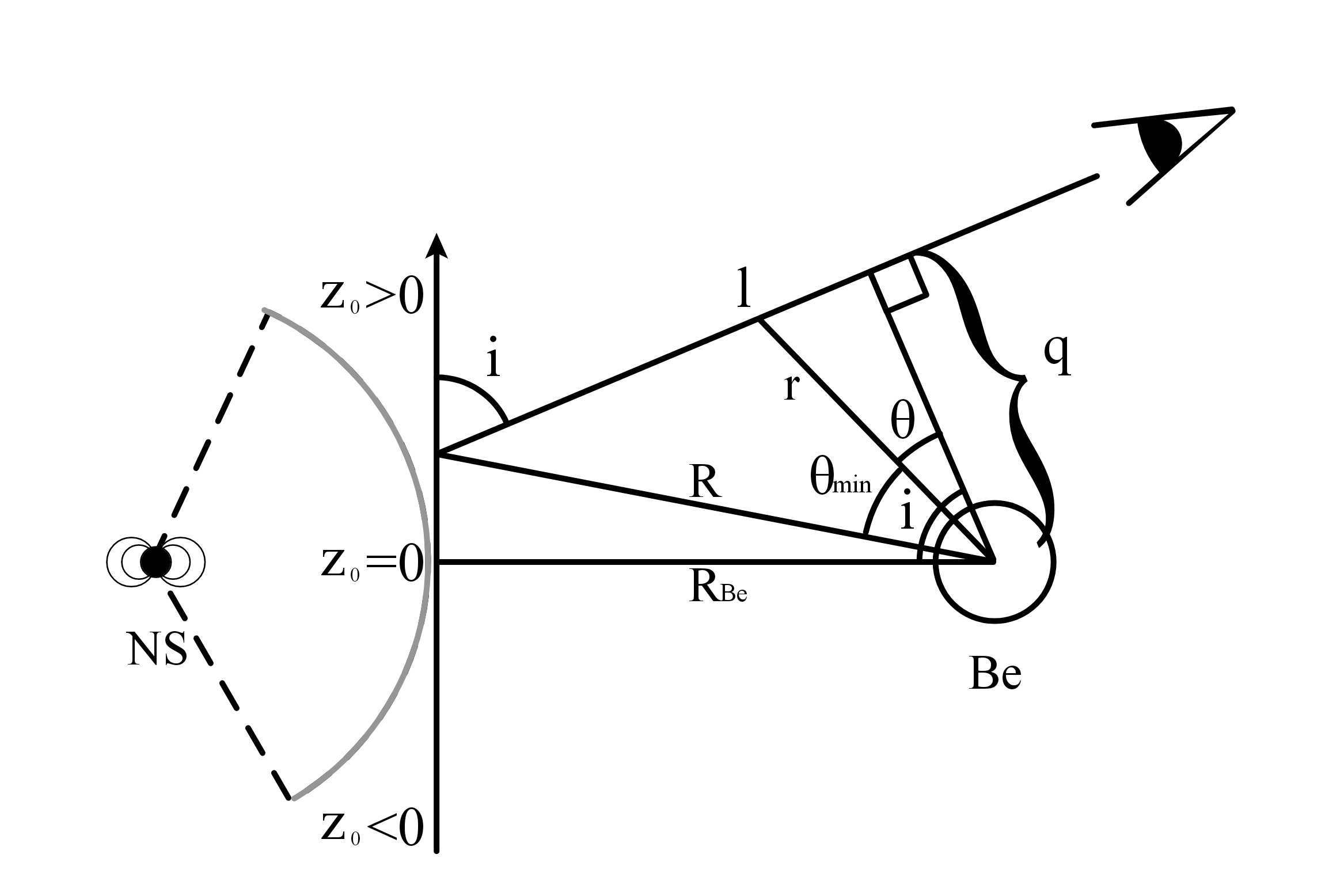}
    \includegraphics[width=0.49\textwidth]{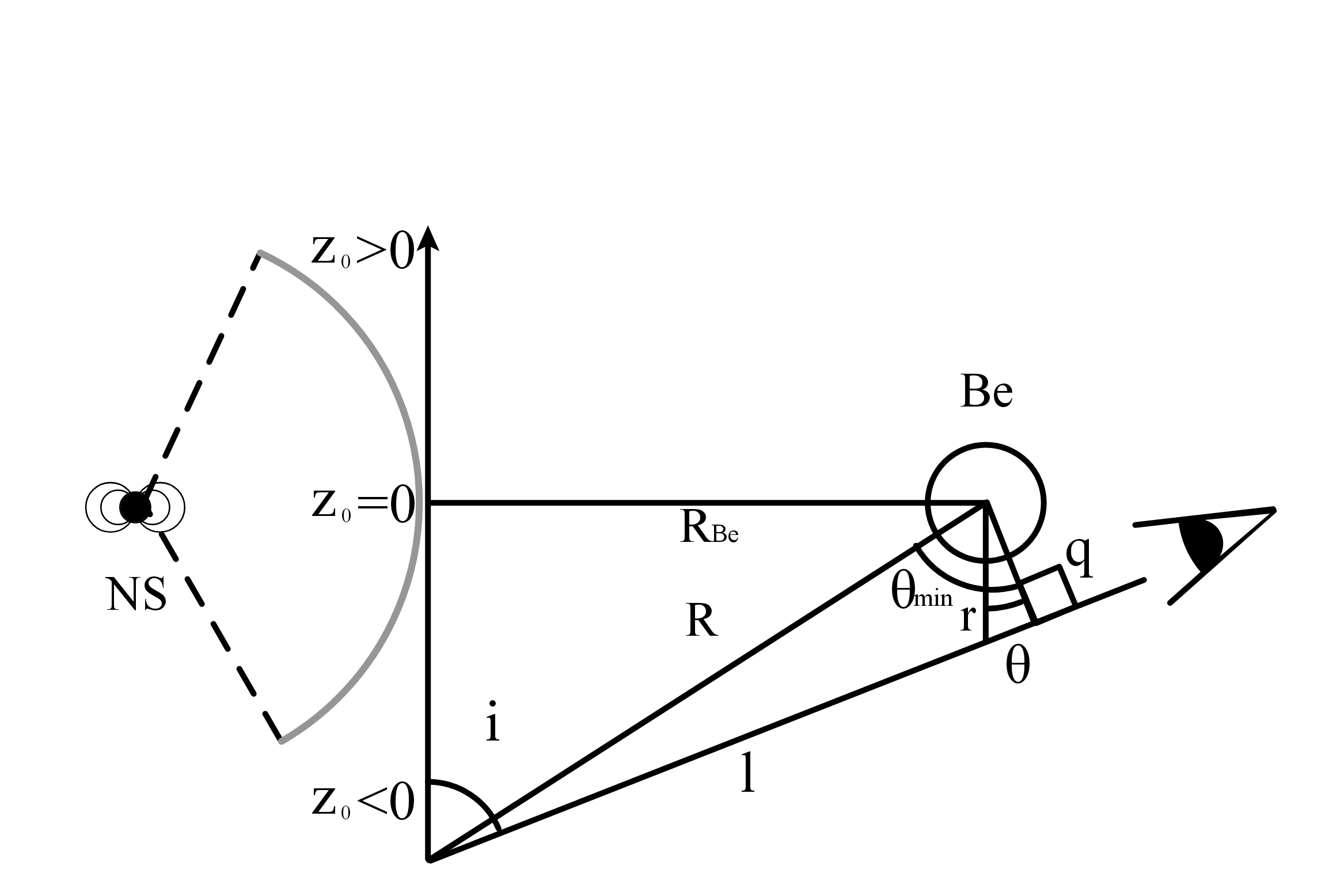}
    \caption{Sketch of the geometry assumed for the free-free optical depth calculation. For an explanation of symbols, see text.}
    \label{fig:ff-geo}
\end{figure}

For a general path (not going through the star) the optical depth for free-free absorption by the stellar wind can be calculated as \citep{1975MNRAS.170...41W}
\begin{equation}
\label{eq:tau_ff}
\tau_{\rm \nu}^{\rm ff} = K(\nu,T) \int_{0}^{+\infty} {\rm d}l \, \frac{A^2}{r^4}
\end{equation}
where $A$ and $K(\nu,T)$ are defined in equations~(\ref{eq:A}) and (\ref{eq:Kv}), respectively. Ignoring the curvature of the shock, we compute the optical depth for paths above and below the apex as shown in Fig.~\ref{fig:ff-geo}, as well as through the apex itself. 

For a path with $z_{\rm o}=0$, equation (\ref{eq:tau_ff}) becomes
\begin{equation}
    \label{eq:path1}
    \tau_{\rm \nu}^{\rm ff}= K(\nu,T) \frac{A^2}{q^3}\int_{-i}^{+\pi/2} {\rm d}\theta \cos^2\theta,
\end{equation}
after changing the variable of integration from $l$ to $\theta$, using that $l=R_{\rm Be}\sin i +q\tan \theta$, ${\rm d} l= (q / \cos^2 \theta) {\rm d}\theta$, and $q=R_{\rm Be}\cos i =r\cos \theta$. 

The above equation can be generalized for paths with $z_{\rm o} \neq 0$ as

\begin{equation}
   \tau_{\rm \nu}^{\rm ff}= K(\nu,T) \frac{A^2}{q^3}\int_{-\theta{\rm min}}^{+\pi/2} {\rm d}\theta \cos^2 \theta
\end{equation} 
where we used $l= \sqrt{z_{\rm o}^2+R_{\rm Be}^2} \sin\theta_{\rm min}+q \tan \theta$. Here, 
\begin{eqnarray}
   \theta_{\rm min} & = & i-\tan^{-1}\left(\frac{|z_{\rm o}|}{R_{\rm Be}}\right) \, {\rm and} \\ 
    q & = & \sqrt{z_{\rm o}^2+R_{\rm Be}^2}\cos\theta_{\rm min}.
\end{eqnarray}

\section{Analytical derivation of the $L_R-L_X$ relation}\label{app:LrLx}
In this section, we derive the scaling relation between the radio and X-ray luminosities expected in our model whenever the radio band is dominated by optically thin synchrotron emission produced in the wind collision region. We also focus on the slow-cooling regime for radio-emitting electrons.

Let $L_R$ be the optically thin synchrotron luminosity at frequency $\nu_R$, i.e. 
\begin{equation}
L_R \equiv P_{\rm syn}(\nu_R) = \frac{4}{3}\sigma_{\rm T} c U_{\rm B} \gamma_R^2 \nu_R^{-1} \left(n_{\gamma_R} \gamma_R \right) V_{\rm em} 
\label{eq:LR}
\end{equation}
where $V_{\rm em} = 4 \pi f_{\rm V} R_{\rm sh}^3$ is the emitting volume of the wind collision region with $f_{\rm V}\approx 1/4$,  
$\gamma_R = \left(2 \pi m_{\rm e} c \nu_R/ (e (8\pi U_{\rm B})^{1/2}) \right)^{1/2}$ is the Lorentz factor of synchrotron emitting electrons at $\nu_R$, and $\left(n_{\gamma_R} \gamma_R \right)$ is the number density of electrons radiating at $\nu_R$. For the derivation of the equation we assume that each electron emits at the characteristic synchrotron frequency. 

Henceforth, we focus on cases with $\nu_R < \nu_{\rm c}$, where $\nu_{\rm c}$ is the characteristic synchrotron cooling frequency. Using also equations (\ref{eq:Rm}) - (\ref{eq:UBe}) and noting that $R_{\rm sh}=\min(d_{\rm NS}, d_{\rm Be})$ the radio luminosity at $\nu_R$ (see equation~\ref{eq:LR}) is written as 
\begin{eqnarray}
L_R & \simeq &   C_1(p) f(p) \,L_{X}^{\frac{12}{7}} \nu_R^{\frac{-p+1}{2}} \left(\dot{M}_{\rm Be} V_{\rm Be}^{\infty} \right)^{\frac{p-1}{4}} \epsilon_{\rm B}^{\frac{p+1}{4}} \epsilon_{\rm e}  \\ \nonumber 
& &  \epsilon^{-\frac{12}{7}} (\eta_{\rm out} \chi)^{\frac{3}{2}} \xi^{-\frac{3}{4}} \mu_{\rm NS}^{-\frac{3}{7}}  D^{\frac{1-p}{2}} 
\label{eq:LrLx-gen-1}
\end{eqnarray}
for $\eta \gg 1$. Here, 
\begin{eqnarray}
    f(p) = \left \{ \begin{tabular}{cc}
                    $(p-2)\gamma_{\min}^{p-2}$, &  $p>2$ and $\gamma_{\max}\gg \gamma_{\min}$ \\  \\
                    $\ln^{-1}\left(\frac{\gamma_{\max}} {\gamma_{\min}}\right)$, & $p=2$
                    \end{tabular}
    \right. 
\end{eqnarray}
and 
\begin{eqnarray}
C_1(p) = \frac{2^{\frac{3}{28}} \sigma_{\rm T} e^{\frac{-3+p}{2}}}{24 \ c^{\frac{24}{7}}}  \left(\frac{9}{4}\right)^{\frac{5+p}{4}} \left(2\pi m_{\rm e}c\right)^{\frac{-p+1}{2}} (GM_{\rm NS})^{\frac{6}{7}} f_{\rm V}
\end{eqnarray}
Hence, for $\eta \gg 1$  (or $L_X \ll L_{X, \rm br}$, see equation~\ref{eq:Lxcr}) and $p=2$ the predicted relation is 
\begin{eqnarray}
L_R &\simeq& 10^{18}\, L_{X, 38}^{\frac{12}{7}} \nu_{R,10}^{-\frac{1}{2}} \left(\dot{M}_{\rm Be,-9} V_{\rm Be, 3}^{\infty} \right)^{\frac{1}{4}} \epsilon_{\rm B, -0.3 }^{\frac{3}{4}} \epsilon_{\rm e, -0.3}   \nonumber \\
& &  \epsilon_{-1}^{-\frac{12}{7}} \ (\eta_{\rm out, -2} \,  \chi_0)^{\frac{3}{2}} \ \xi_{-0.3}^{-\frac{3}{4}}\  \mu_{\rm NS, 30}^{-\frac{3}{7}} \nonumber \\
& & D_{13}^{-\frac{1}{2}} \ \ln^{-1}\left(\frac{\gamma_{\max}}{\gamma_{\min}}\right) {\rm erg \, s^{-1} \, Hz^{-1}}  
\label{eq:LrLx-1}
\end{eqnarray}
where we introduced the notation $q_x \equiv q/10^x$ in cgs units, except for the Be mass loss rate that is in units of $M_{\odot}/{\rm yr}$ and the wind velocity that is in units of km/s.

Similarly, we derive the expression for $\eta \ll 1$, which reads
\begin{eqnarray}
L_R &\simeq & C_2(p) f(p)\, L_{X}^{\frac{2(p-1)}{7}} \nu_R^{\frac{-p+1}{2}} \left(\dot{M}_{\rm Be} V_{\rm Be}^{\infty} \right)^{\frac{3}{2}} \epsilon_{\rm B}^{\frac{p+1}{4}}  \\ \nonumber 
& &  \epsilon_{\rm e} \  \epsilon^{\frac{2(1-p)}{7}} (\eta_{\rm out} \chi)^{\frac{p-1}{4}} \xi^{\frac{1-p}{8}} \mu_{\rm NS}^{\frac{1-p}{14}}  D^{\frac{1-p}{2}} 
\label{eq:LrLx-gen-2}
\end{eqnarray}
with $C_2$ given by
\begin{eqnarray}
C_2(p) = \frac{2^{\frac{p-1}{56}} \sigma_{\rm T} e^{\frac{-3+p}{2}}}{24 \ c^{\frac{4(p-1)}{7}}}  \left(\frac{9}{4}\right)^{\frac{5+p}{4}} \left(2\pi m_{\rm e}c\right)^{\frac{-p+1}{2}}    (GM_{\rm NS})^{\frac{p-1}{7}} f_{\rm V}
\end{eqnarray}
Hence, for $\eta \ll 1$ (or $L_X \gg L_{X, \rm br}$) and $p=2$, the expected radio-X-ray luminosity relation reads
\begin{eqnarray}
L_R &\simeq& 3.5\times 10^{17}\, L_{X, 38}^{\frac{2}{7}} \nu_{R,10}^{-\frac{1}{2}} \left(\dot{M}_{\rm Be,-9} V_{\rm Be, 3}^{\infty} \right)^{\frac{3}{2}} \epsilon_{\rm B, -0.3 }^{\frac{3}{4}} \epsilon_{\rm e, -0.3}   \nonumber \\ 
& & \epsilon_{-1}^{-\frac{2}{7}} \ (\eta_{\rm out, -2} \,  \chi_0)^{\frac{1}{4}} \ \xi_{-0.3}^{-\frac{1}{8}}\  \mu_{\rm NS, 30}^{-\frac{1}{14}} \nonumber \\ 
& & D_{13}^{-\frac{1}{2}} \  \ln^{-1}\left(\frac{\gamma_{\max}}{\gamma_{\min}}\right) {\rm erg \, s^{-1} \, Hz^{-1}}   
\label{eq:LrLx-2}
\end{eqnarray}

\section{Corner plots}\label{app:corner}

\begin{figure*}
    \centering
    \includegraphics[width=\textwidth]{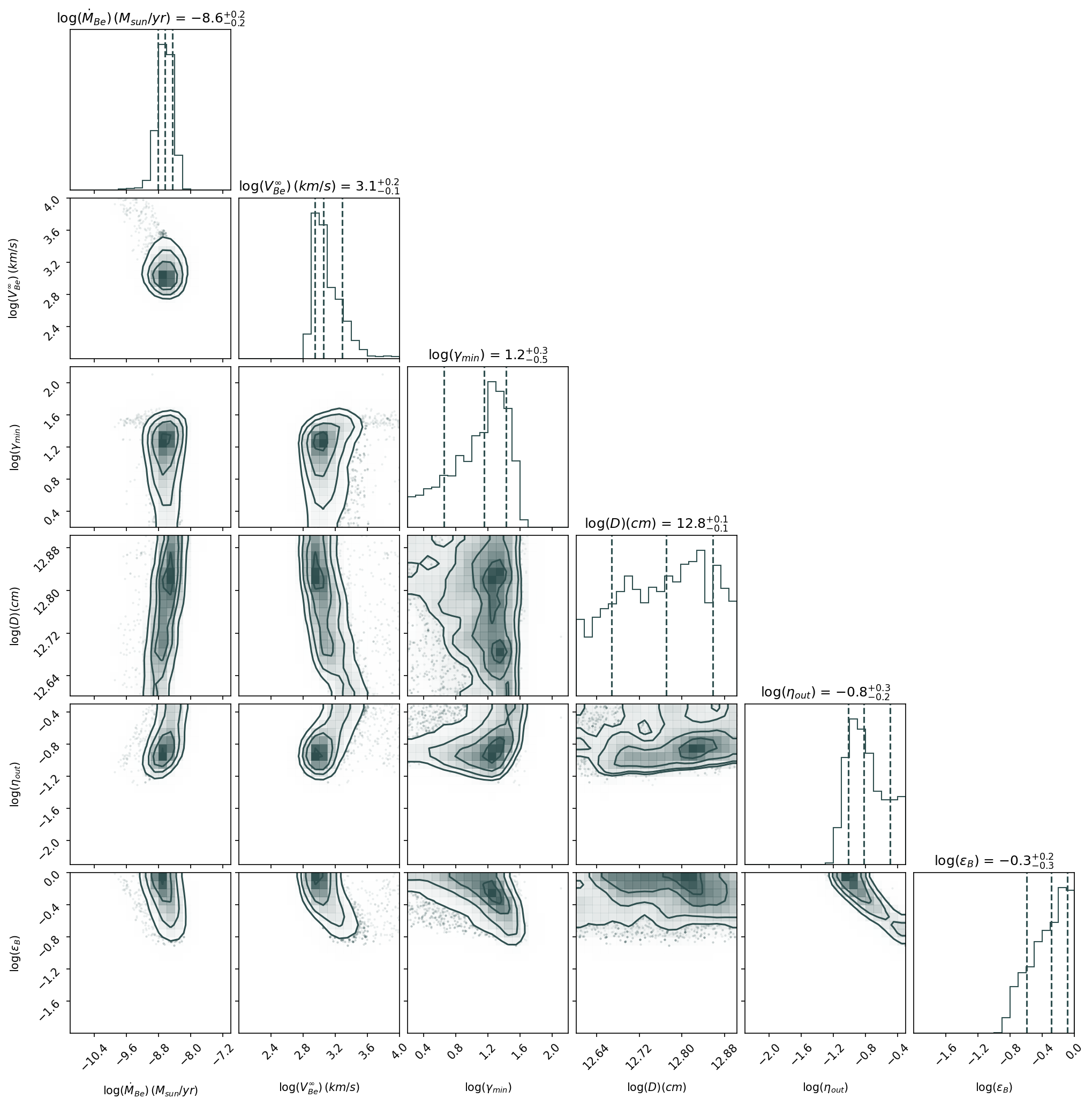}
    \caption{Corner plot of the parameter pairs and the marginal posterior distributions for epoch A. Dashed vertical lines indicate the median and the 68 per cent range for each parameter.}
    \label{fig:corner-epA}
\end{figure*}

\begin{figure*}
    \centering
    \includegraphics[width=\textwidth]{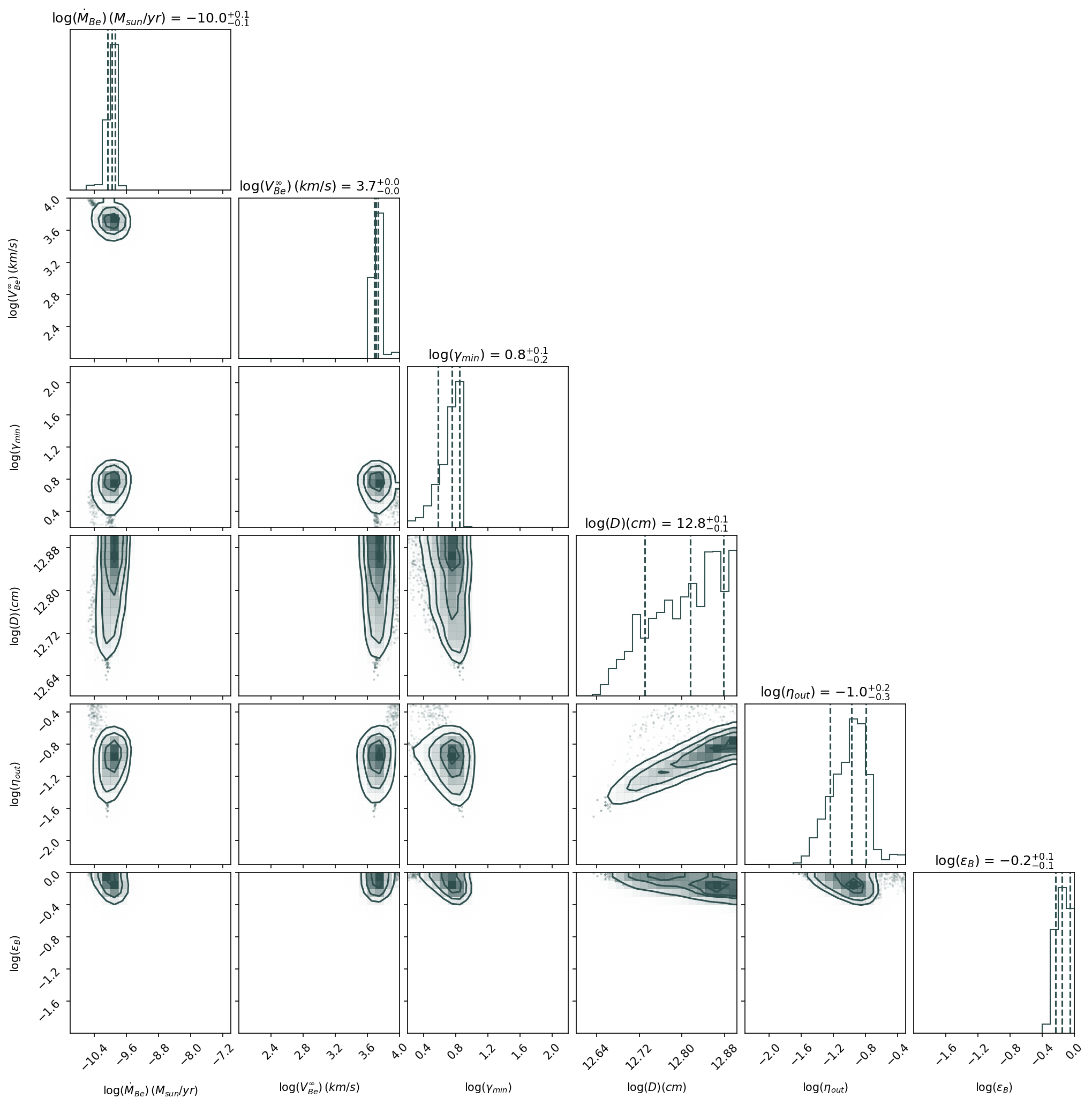}
    \caption{Corner plot of the parameter pairs and the marginal posterior distributions for epoch B. Dashed vertical lines indicate the median and the 68 per cent range for each parameter.}
    \label{fig:corner-epB}
\end{figure*}




\bsp	
\label{lastpage}
\end{document}